\documentclass[lettersize,journal]{IEEEtran}
\usepackage{amsmath,amsfonts}
\usepackage{algorithm}
\usepackage{algpseudocode}
\usepackage{array}
\usepackage[caption=false,font=normalsize,labelfont=sf,textfont=sf]{subfig}
\usepackage{textcomp}
\usepackage{stfloats}
\usepackage{url}
\usepackage{verbatim}
\usepackage{graphicx}
\usepackage{cite}
\usepackage{diagbox}
\usepackage{multirow}
\usepackage{lipsum}
\usepackage{amssymb, amsthm}
\usepackage{booktabs}
\usepackage{xcolor}
\usepackage{amsmath}
\usepackage{arydshln}

\makeatletter
\def\adl@drawiv#1#2#3{%
        \hskip.5\tabcolsep
        \xleaders#3{#2.5\@tempdimb #1{1}#2.5\@tempdimb}%
                #2\z@ plus1fil minus1fil\relax
        \hskip.5\tabcolsep}
\newcommand{\cdashlinelr}[1]{%
  \noalign{\vskip\aboverulesep
           \global\let\@dashdrawstore\adl@draw
           \global\let\adl@draw\adl@drawiv}
  \cdashline{#1}
  \noalign{\global\let\adl@draw\@dashdrawstore
           \vskip\belowrulesep}}
\makeatother

\newcommand{\Rbb}{\mathbb{R}}

\newcommand{\Ebb}{\mathbb{E}}

\newcommand{\Cbb}{\mathbb{C}}

\newcommand{\Tcal}{\mathcal{T}}
\newcommand{\Acal}{\mathcal{A}}

\newcommand{\Ncal}{\mathcal{N}}

\newcommand{\Dcal}{\mathcal{D}}

\def\*#1{\mathbf{#1}}
\newcommand\mbfsf[1]{\boldsymbol{\mathsf{#1}}}

\newcommand{\btheta}{\boldsymbol{\theta}}
\newcommand{\bsigma}{\boldsymbol{\Sigma}}
\newcommand{\bpsi}{\boldsymbol{\psi}}
\newcommand{\bphi}{\boldsymbol{\phi}}
\newcommand{\bmu}{\boldsymbol{\mu}}

\begin{document}

%%%%%%%%%%%%%%%%%%%%%%%%%%%%%%%%%%%%%%%%%%%%%%%%%%%%%%%%%%%%%%%%%%%%%%%%%%%%%%%%%%%%
%%%%%%%%%%% TITLE
%%%%%%%%%%%%%%%%%%%%%%%%%%%%%%%%%%%%%%%%%%%%%%%%%%%%%%%%%%%%%%%%%%%%%%%%%%%%%%%%%%%%
% \title{Integrating Generative Models and Bayesian Neural Networks for Uncertainty Quantification in Computational Imaging}

% \title{Conformalized Bayesian Imaging: A Framework for Rigorous Uncertainty Quantification in Imaging Inverse Problems}

% \title{Conformalized Bayesian Imaging: An Uncertainty Quantification Framework for Computational Imaging}

\title{Conformalized Generative Bayesian Imaging: \\ An Uncertainty Quantification Framework for Computational Imaging}

%%%%%%%%%%%%%%%%%%%%%%%%%%%%%%%%%%%%%%%%%%%%%%%%%%%%%%%%%%%%%%%%%%%%%%%%%%%%%%%%%%%%
%%%%%%%%%%% AUTHOR INFO AND THANKS
%%%%%%%%%%%%%%%%%%%%%%%%%%%%%%%%%%%%%%%%%%%%%%%%%%%%%%%%%%%%%%%%%%%%%%%%%%%%%%%%%%%%
\author{Canberk Ekmekci,~\IEEEmembership{Graduate Student Member,~IEEE}, and Mujdat Cetin,~\IEEEmembership{Fellow,~IEEE}
        % <-this % stops a space
\thanks{This work was partially supported by the National Science Foundation (NSF) under Grant CCF-1934962.}
\thanks{Canberk Ekmekci is with the Department of Electrical and Computer Engineering, University of Rochester, Rochester, 
NY 14627 USA (e-mail: cekmekci@ur.rochester.edu).}
\thanks{Mujdat Cetin is with the Department of Electrical and Computer Engineering and Goergen Institute for Data Science \& AI, University of Rochester, Rochester, NY 14627 USA (e-mail: mujdat.cetin@rochester.edu).}
}

% Uncomment the following for double coulmn IEEE paper


\maketitle

\bstctlcite{IEEEexample:BSTcontrol} 

%%%%%%%%%%%%%%%%%%%%%%%%%%%%%%%%%%%%%%%%%%%%%%%%%%%%%%%%%%%%%%%%%%%%%%%%
%%%%%%%%%%% ABSTRACT
%%%%%%%%%%%%%%%%%%%%%%%%%%%%%%%%%%%%%%%%%%%%%%%%%%%%%%%%%%%%%%%%%%%%%%%%
\begin{abstract}
Uncertainty quantification plays an important role in achieving trustworthy and reliable learning-based computational imaging. Recent advances in generative modeling and Bayesian neural networks have enabled the development of uncertainty-aware image reconstruction methods. Current generative model-based methods seek to quantify the inherent (aleatoric) uncertainty on the underlying image for given measurements by learning to sample from the posterior distribution of the underlying image. On the other hand, Bayesian neural network-based approaches aim to quantify the model (epistemic) uncertainty on the parameters of a deep neural network-based reconstruction method by approximating the posterior distribution of those parameters. Unfortunately, an ongoing need for an inversion method that can jointly quantify complex aleatoric uncertainty and epistemic uncertainty patterns still persists. In this paper, we present a scalable framework that can quantify both aleatoric and epistemic uncertainties. The proposed framework accepts an existing generative model-based posterior sampling method as an input and introduces an epistemic uncertainty quantification capability through Bayesian neural networks with latent variables and deep ensembling. Furthermore, by leveraging the conformal prediction methodology, the proposed framework can be easily calibrated to ensure rigorous uncertainty quantification. We evaluated the proposed framework on magnetic resonance imaging, computed tomography, and image inpainting problems and showed that the epistemic and aleatoric uncertainty estimates produced by the proposed framework display the characteristic features of true epistemic and aleatoric uncertainties. Furthermore, our results demonstrated that the use of conformal prediction on top of the proposed framework enables marginal coverage guarantees consistent with frequentist principles. 
\end{abstract}

%%%%%%%%%%%%%%%%%%%%%%%%%%%%%%%%%%%%%%%%%%%%%%%%%%%%%%%%%%%%%%%%%%%%%%%%
%%%%%%%%%%% KEYWORDS
%%%%%%%%%%%%%%%%%%%%%%%%%%%%%%%%%%%%%%%%%%%%%%%%%%%%%%%%%%%%%%%%%%%%%%%%

%%%%%%%%%%%%%%%%%%%%%%%%%%%%%%%%%%%%%%%%%%%%%%%%%%%%%%%%%%%%%%%%%%%%%%%%
%%%%%%%%%%% INTRODUCTION 
%%%%%%%%%%%%%%%%%%%%%%%%%%%%%%%%%%%%%%%%%%%%%%%%%%%%%%%%%%%%%%%%%%%%%%%%
\section{Introduction}
\label{sec:introduction}

\IEEEPARstart{T}{his} paper focuses on two main types of uncertainties arising in computational imaging problems, namely the \emph{aleatoric} uncertainty~\cite{Kiureghian2009AleatoricEpistemic,Hullermeier2021AleatoricEpistemic} and the \emph{epistemic} uncertainty~\cite{Kiureghian2009AleatoricEpistemic, Hullermeier2021AleatoricEpistemic}. For a given imaging inverse problem, aleatoric uncertainty refers to the inherent randomness on the underlying image for a given set of measurements. This type of uncertainty arises from the ill-posed nature of the problem and would remain even with the optimal reconstruction method and unlimited training data. It cannot be reduced without modifying the imaging setup or changing the formulation of the inverse problem. In contrast, epistemic uncertainty refers to the uncertainty arising from incomplete knowledge about a statistical prediction model. Unlike aleatoric uncertainty, this type of uncertainty can be reduced with more data or additional knowledge about the model. In the context of imaging, epistemic uncertainty often appears as the uncertainty on the adjustable parameters of an image reconstruction method used to solve the inverse problem. For a given deep learning-based image reconstruction method, epistemic uncertainty corresponds to the uncertainty on the parameters of the underlying deep neural network~\cite{Blundell2015BayesianNN}, which arises due to a lack of training data in the neighborhood of a test sample. Thus, in contrast to aleatoric uncertainty, the epistemic uncertainty is, in principle, reducible by collecting more training examples~\cite[Section 2]{Hullermeier2021AleatoricEpistemic}.

Designing deep learning-based image reconstruction methods that are capable of quantifying the aleatoric uncertainty and the epistemic uncertainty is crucial for identifying possible solutions of an imaging inverse problem and determining how uncertain the deep learning-based image reconstruction method is about those solutions. In the computational imaging literature, two classes of approaches have been followed to quantify these uncertainties: generative model-based posterior sampling methods and Bayesian neural network-based image reconstruction techniques.

From a Bayesian perspective, the aleatoric uncertainty can be represented by the posterior distribution of the underlying image given measurements. Thanks to the advancements in deep generative modeling (see \cite{BondTaylor2021GenerativeSurvey} for a survey), several generative model-based posterior sampling methods, e.g., \cite{Ardizzone2019GenerativeModelPosteriorSampling, Bohm2019VAEGenerativeModelPosteriorSampling, Bohra2022GenerativeModelPosteriorSampling, Choi2021GenerativeModelPosteriorSampling, Liu2022GenerativeModelPosteriorSampling, Meng2022GenerativeModelPosteriorSampling, Song2022GenerativeModelPosteriorSampling, Song2023GenerativeModelPosteriorSampling, Zhao2023GenerativeModelPosteriorSampling, Chung2023DiffusionPosteriorSampling, Kawar2022GenerativeModelPosteriorSampling, Ramzi2020GenerativeModelPosteriorSampling, Whang2021GenerativeModelPosteriorSampling, Chung2023GenerativeModelPosteriorSampling, McCann2023GenerativeModelPosteriorSampling, Dasgupta2021GenerativeModelPosteriorSampling, Tonolini2020GenerativeModelPosteriorSampling, Adler2019CWGAN, Sun2021DPI}, have been proposed to quantify the aleatoric uncertainty appearing in imaging inverse problems. 
% make changes below
These methods have leveraged various deep generative models such as variational autoencoders~\cite{Kingma2013VAE, Sohn2015CVAEPaper}, generative adversarial networks~\cite{Mirza2014ConditionalGAN, Goodfellow2014GAN}, flow-based generative models~\cite{Dinh2015NormalizingFlow,Dinh2017RealNVPNormalizingFlow}, and diffusion models~\cite{Dickstein2015DiffusionModels, Ho2020DDPM, Song2021StochasticDifferentialEquations, Song2019ScoreBasedModels} to learn the posterior distribution of the underlying image from data, i.e., to quantify the aleatoric uncertainty arising in imaging inverse problems. They have found use in several imaging problems such as super-resolution~\cite{Song2023GenerativeModelPosteriorSampling, Kawar2022GenerativeModelPosteriorSampling, Tonolini2020GenerativeModelPosteriorSampling, Whang2021GenerativeModelPosteriorSampling, Chung2023DiffusionPosteriorSampling}, inpainting~\cite{Song2023GenerativeModelPosteriorSampling, Kawar2022GenerativeModelPosteriorSampling, Tonolini2020GenerativeModelPosteriorSampling,Whang2021GenerativeModelPosteriorSampling, McCann2023GenerativeModelPosteriorSampling, Chung2023DiffusionPosteriorSampling}, JPEG restoration~\cite{Song2023GenerativeModelPosteriorSampling}, deblurring~\cite{Kawar2022GenerativeModelPosteriorSampling, Tonolini2020GenerativeModelPosteriorSampling, Chung2023DiffusionPosteriorSampling}, phaseless holographic imaging~\cite{Tonolini2020GenerativeModelPosteriorSampling}, imaging through scattering media~\cite{Tonolini2020GenerativeModelPosteriorSampling}, compressed sensing~\cite{Whang2021GenerativeModelPosteriorSampling}, blind image deblurring~\cite{Chung2023GenerativeModelPosteriorSampling}, imaging through turbulence~\cite{Chung2023GenerativeModelPosteriorSampling}, magnetic resonance imaging~\cite{Ramzi2020GenerativeModelPosteriorSampling, Song2022GenerativeModelPosteriorSampling}, phase retrieval~\cite{Bohra2022GenerativeModelPosteriorSampling}, optical diffraction tomography~\cite{Bohra2022GenerativeModelPosteriorSampling}, nonlinear Fourier magnitude retrieval~\cite{McCann2023GenerativeModelPosteriorSampling}, and limited-angle computed tomography~\cite{Liu2022GenerativeModelPosteriorSampling, Song2022GenerativeModelPosteriorSampling, Adler2019CWGAN}, demonstrating the aleatoric uncertainty characterization capability of generative model-based posterior sampling methods.

Similarly, Bayesian neural network~\cite{Neal1995BayesianNN}-based image reconstruction methods also take a Bayesian approach; however, they aim to capture the epistemic uncertainty on the parameters of a deep learning-based image reconstruction method by learning the posterior distribution of the parameters given a training dataset containing target image-measurement pairs. Unfortunately, calculating the exact posterior distribution of the parameters is mathematically intractable due to the deep non-linear structure of modern deep neural networks. Therefore, various techniques have been developed to tackle this challenge, such as specialized variational inference techniques~\cite{Gal2016MCDropout, Blundell2015BayesianNN, Graves2011VariationalInferenceBNNs}, scalable Markov Chain Monte Carlo methods~\cite{Chen14MCMCBNNs, Ma2015MCMCBNNs, Welling2011MCMCBNNs, Zhang2020MCMCBNNs}, and practitioner-friendly deep ensembling methods~\cite{Lakshminarayanan2017ensembling} (see \cite{Jospin2022BNNTutorial} for a comprehensive overview). As a result of these advancements, several Bayesian neural network-based image reconstruction methods have been developed in the literature (e.g., \cite{Schlemper2018BNNUNetAndUnrolledForMRI, Xue2019BNNImageReconstruction, Shang2021BNNImageReconstruction, Tanno2019BNNImageReconstruction, Ekmekci2021BNNImageReconstruction, Ekmekci2022BNNImageReconstruction, Cochrane2022BNNImageReconstruction, Hoffmann2021BNNImageReconstruction, Siahkoohi2020BNNImageReconstruction}) focusing on various imaging problems, including but not limited to, MRI super-resolution~\cite{Tanno2019BNNImageReconstruction}, phase imaging~\cite{Xue2019BNNImageReconstruction}, seismic imaging~\cite{Siahkoohi2020BNNImageReconstruction}, computational optical form measurements~\cite{Hoffmann2021BNNImageReconstruction}, single-pixel imaging~\cite{Shang2021BNNImageReconstruction}, and imaging through scattering media~\cite{Cochrane2022BNNImageReconstruction}, illustrating the epistemic uncertainty characterization capability of Bayesian neural network-based image reconstruction methods. Furthermore, alongside problem-specific methodological developments, more general Bayesian neural network-based image reconstruction methods, e.g., \cite{Ekmekci2021BNNImageReconstruction, Ekmekci2022BNNImageReconstruction}, have been developed to enable the use of Bayesian neural networks for a wide variety of imaging applications. 
% Ardindan en sonunda ornek verirken sunlari kullan: such as deep learning-based post-processing methods (e.g., \cite{Jin2017FBPConvNet}), deep unrolling methods~\cite{Gregor2010Unfolding} (see \cite{Monga2021Unrolling} for a comprehensive survey), and Plug-and-Play methods~\cite{Venkatakrishnan2013PnP} (see \cite{Kamilov2023PnPSurvey} for a recent survey). 
% to capture the epistemic uncertainty for different deep learning-based image reconstruction methodologies such as post-processing-type methods, deep unrolling methods~\cite{Gregor2010Unfolding, Monga2021Unrolling}, and Plug-and-Play priors~\cite{Venkatakrishnan2013PnP}. 

Although generative model-based posterior sampling methods and Bayesian neural network-based image reconstruction methods are capable of quantifying the aleatoric uncertainty and the epistemic uncertainty respectively, they do not provide simultaneous aleatoric and epistemic uncertainty estimates. Specifically, generative model-based posterior sampling methods are capable of quantifying complex aleatoric uncertainty patterns; however, they rely on a single set of parameters at the inference stage. Thus, they neither capture the epistemic uncertainty on the parameters of the generative models nor provide any information about how uncertain the generative model is about the generated samples. On the contrary, Bayesian neural network-based image reconstruction methods can quantify the epistemic uncertainty on the parameters while being incapable of quantifying the aleatoric uncertainty. Certain modifications on Bayesian neural network-based image reconstruction methods can render them capable of capturing the aleatoric uncertainty, e.g., \cite{Schlemper2018BNNUNetAndUnrolledForMRI, Tanno2019BNNImageReconstruction, Xue2019BNNImageReconstruction, Shang2021BNNImageReconstruction, Ekmekci2022BNNImageReconstruction, Ekmekci2021BNNImageReconstruction}, but those modifications make simplifying assumptions about the form of the aleatoric uncertainty, and those assumptions do not necessarily hold for all imaging inverse problems, especially for the ones where the posterior distribution of the underlying image tends to be highly multimodal.

To address these limitations, this paper presents a framework that can quantify both epistemic and complex aleatoric uncertainty patterns, presenting a comprehensive solution to the shortcomings of the existing generative model-based posterior sampling methods and Bayesian neural network-based image reconstruction methods. The proposed framework accepts an existing generative model-based posterior sampling method as an input and enhances it by introducing epistemic uncertainty capability via Bayesian neural networks with latent variables (BNN+LV)~\cite{Depeweg2017BNNLV, Depeweg2018BNNLV}. Furthermore, we suggest that by incorporating conformal prediction techniques~\cite{Vovk1999ConformalPrediction, Volk2005ConformalPredictionBook, Lei2013ConformalPrediction, Lei2015ConformalPrediction, Angelopoulos2023ConformalPrediction}, the proposed framework can be conformalized to ensure rigorous uncertainty estimation with reliable marginal coverage guarantees. We evaluate the proposed framework on various image reconstruction and restoration problems, namely computed tomography (CT), magnetic resonance imaging (MRI), and image inpainting. We analyze the behavior of the aleatoric and epistemic uncertainty estimates provided by the proposed framework under a variety of experimental conditions. We demonstrate that the epistemic and aleatoric uncertainty estimates produced by the proposed framework display the characteristics of true uncertainties. Moreover, we show that the conformalized version of the proposed framework is capable of providing reliable uncertainty estimates, ensuring that the resulting conformal prediction set satisfies the marginal coverage property.

\subsection{Contributions}
\label{ssec:contributions}
The contributions of this paper are three-fold: 
\begin{itemize}
    \item We propose an image reconstruction framework that has the ability to quantify both epistemic and aleatoric uncertainties by enhancing the uncertainty characterization capability of a given  generative model-based posterior sampling method with Bayesian neural networks with latent variables. Thanks to the utilization of deep ensembling~\cite{Lakshminarayanan2017ensembling} during the training stage, the proposed framework is versatile for various imaging applications.
    \item We reveal the connection between generative model-based posterior sampling methods and Bayesian neural network-based image reconstruction methods by comparing each of them individually with the proposed framework, addressing a conceptual gap that has not been explored in the existing computational imaging literature.
    \item We provide a calibration guide for the proposed framework using conformal prediction and demonstrate that conformalization yields prediction sets that satisfy the marginal coverage property. This confirms the proposed framework's ability to deliver rigorous uncertainty estimates.
\end{itemize}

\subsection{Comparison with Prior Work}
\label{ssec:related-work-comparison}
The main difference between the Bayesian neural network-based image reconstruction methods and the proposed framework is that Bayesian neural network-based approaches quantify only the epistemic uncertainty while the proposed framework captures both the aleatoric uncertainty and the epistemic uncertainty. However, it is worth noting that there are certain Bayesian neural network-based image reconstruction methods, e.g., \cite{Schlemper2018BNNUNetAndUnrolledForMRI, Tanno2019BNNImageReconstruction, Xue2019BNNImageReconstruction, Shang2021BNNImageReconstruction, Ekmekci2022BNNImageReconstruction, Ekmekci2021BNNImageReconstruction}, aiming to characterize the aleatoric uncertainty as well by making relatively restrictive assumptions on the form of the aleatoric uncertainty. The key difference between such approaches and the proposed framework lies in the use of a latent variable in problem formulation. As we will show later in Section \ref{ssec:proposed-method-comparison-with-bnn}, such Bayesian neural network-based image reconstruction methods model the aleatoric uncertainty as noise, e.g., additive Gaussian noise, while the proposed framework models the aleatoric uncertainty in a more complex way by using a latent variable in the problem formulation.

Turning to generative model-based posterior sampling methods, it becomes apparent that the fundamental difference between the generative model-based posterior exploration methods and the proposed framework is that generative model-based approaches are not capable of quantifying epistemic uncertainty since they only use point estimates of the parameters of generators. Conversely, the proposed framework has the ability to capture the epistemic uncertainty by using an ensemble of estimates of the parameters through deep ensembling~\cite{Lakshminarayanan2017ensembling}.

It is worth mentioning that the preliminary version of this work has appeared as a conference paper in \cite{Ekmekci2023PreliminaryBNNLV}. This manuscript extends the ideas presented in the preliminary work in several significant ways. While the preliminary version of this work has introduced the building blocks of the proposed framework, it does not explore the conceptual and mathematical connection between the proposed framework and other uncertainty quantifying frameworks, namely generative model-based posterior sampling methods and Bayesian neural network-based image reconstruction methods. This work provides such discussion in Section \ref{ssec:proposed-method-comparison-with-gm} and Section \ref{ssec:proposed-method-comparison-with-bnn}. Additionally, the issues of model bias and miscalibration have not been addressed in the preliminary version of this work. This paper addresses those issues using the conformal prediction framework in Section \ref{ssec:conformal-prediction}. Moreover, the experiments provided in the preliminary version did not examine certain aspects of the proposed framework that may be important in practice. This paper provides such an analysis through the experiments presented in Section \ref{ssec:experiments-training-dataset-size-epistemic} and Section \ref{ssec:experiments-mc-dropout}. Furthermore, this work expands the scope of the experiments by demonstrating that the proposed framework can be utilized with various generative model-based posterior sampling methods for a variety of imaging inverse problems.

Following the preliminary version of this work, Chan \emph{et al.}~\cite{Chan2024Hyperdiffusionestimatingepistemicaleatoric} proposed the Hyper-Diffusion model, which combines hyper-networks~\cite{Ha2017hypernetworks} and conditional denoising diffusion models~\cite{Ho2020DDPM} together to estimate aleatoric and epistemic uncertainties arising in imaging inverse problems using a single network. While Hyper-Diffusion offers significant advancements in computational efficiency by eliminating the need for deep ensembling, which is employed in this paper and its preliminary version since it does not require any changes on the training procedures of existing generative model-based posterior sampling methods, this paper provides a more general treatment emphasizing that the proposed framework can be used with a broad class of generative models, not only diffusion models, and in principle, it can be used with any appropriate posterior approximation method or any ensembling method designed for Bayesian neural networks. Moreover, this paper not only provides an uncertainty quantification framework for imaging inverse problems but also explores the connection between generative model-based posterior sampling methods and Bayesian neural network-based image reconstruction methods, addressing an existing conceptual gap in the computational imaging literature. Furthermore, this paper experimentally shows that the proposed framework might produce uncalibrated predictions due to the underlying modeling assumptions and then provides a calibration guide based on the conformal prediction algorithm to ensure rigorous predictions and accurate uncertainty estimates.

% \subsection{Organization}
% \label{ssec:organization}
% The structure of the paper is as follows: Section \ref{sec:proposed-method} introduces the proposed framework and investigates the connection between the proposed framework, Bayesian neural network-based image reconstruction methods, and deep generative model-based posterior sampling methods. Section \ref{sec:experiments-and-results} presents the experiments, and Section \ref{sec:discussion} discusses the results. Section \ref{sec:conclusion} concludes the paper.

\subsection{Notation}
\label{ssec:notation}
Throughout this paper, we denote vectors and matrices with boldface type (e.g., $\*x$ and $\*X$). We denote random quantities such as random variables and vectors with serif type-style (e.g., $\mathsf{x}$ and $\mbfsf{x}$). We denote the probability density function of a random vector $\mbfsf{x}$ with $p_{\mbfsf{x}}$ and use the function $p_{\mbfsf{x} | \mbfsf{y}}(\mbfsf{x} | \*y)$ to denote the conditional probability density function of $\mbfsf{x}$ given $\mbfsf{y} = \*y$. We use the notation $\mbfsf{x} \sim \Ncal(\bmu, \bsigma)$ to express that the random vector $\mbfsf{x}$ is a normal random vector with mean $\bmu$ and covariance matrix $\bsigma$. We denote the probability density function of a normal random vector $\mbfsf{x}$ with mean $\bmu$ and covariance matrix $\bsigma \succ 0$ with $\Ncal(\mbfsf{x} | \bmu, \bsigma)$. We use $\Ebb$ to denote the expectation operator. 
% We denote the set of real numbers and the set of complex numbers with $\Rbb$ and $\Cbb$, respectively.

%%%%%%%%%%%%%%%%%%%%%%%%%%%%%%%%%%%%%%%%%%%%%%%%%%%%%%%%%%%%%%%%%%%%%%%%%%%%%%%%%%%%
%%%%%%%%%%% PROPOSED METHOD
%%%%%%%%%%%%%%%%%%%%%%%%%%%%%%%%%%%%%%%%%%%%%%%%%%%%%%%%%%%%%%%%%%%%%%%%%%%%%%%%%%%%
\section{Proposed Framework}
\label{sec:proposed-method}
This section describes the class of inverse problems of interest and states the assumptions made about the problem setup. It presents the proposed framework in detail and explores the connection between the proposed framework, generative model-based posterior sampling methods, and Bayesian neural network-based image reconstruction techniques.

\subsection{Problem Setup}
\label{ssec:problem-setup}
The proposed framework is suitable for addressing a wide range of imaging inverse problems for which the observation models have the following structure:
\begin{equation}
    \*y = \xi \left( \Acal(\*x) \right),
\end{equation}
where $\*y \in \Cbb^M$ is the measurement vector; $\*x \in \Cbb^N$ is the underlying image in a vectorized form; $\Acal : \Cbb^N \to \Cbb^M$ is the deterministic forward operator modeling the transformation applied to the underlying image during the sensing process; and $\xi : \Cbb^M \to \Cbb^M$ is the stochastic operator modeling the noise in the imaging system.

For such inverse problems, throughout the remainder of this paper, we make two main assumptions about the problem setup. First, we assume access to a training dataset $\Dcal$ containing measurement vectors and corresponding reference images. Second, we assume that we already have a generative model-based posterior sampling method at hand that is capable of generating samples from the posterior distribution of the underlying image given measurements $p_{\mbfsf{x} | \mbfsf{y}}(\mbfsf{x} | \*y)$. In the subsequent sections, we denote this generative model-based posterior sampling method by the tuple $(G, p_{\mbfsf{z}}, \Tcal)$. In this notation, $G : \Cbb^M \times \Rbb^Z \to \Cbb^N$ is a conditional generative model generating the samples; $\mbfsf{z} \sim p_{\mbfsf{z}}$ is an $\Rbb^Z$-valued random latent variable; and $\Tcal$ is the training procedure followed by the generative model-based posterior sampling method to train the generative model $G$. These assumptions are often justifiable for a variety of imaging problems, although they may not be plausible for certain imaging problems, especially if obtaining reference images is unfeasible.

\subsection{Proposed Uncertainty Quantification Approach}
\label{ssec:proposed-framework-details}
As mentioned in Section \ref{ssec:related-work-comparison}, the main limitation of generative model-based posterior sampling methods regarding uncertainty quantification is their lack of quantifying the epistemic uncertainty as they only use a point estimate of the parameters of the underlying generative model. The proposed framework addresses this shortcoming by following the principles of the BNN+LV framework~\cite{Depeweg2017BNNLV, Depeweg2018BNNLV} and treating the parameters of the generative model $G$ as random variables.

Since parameters are treated as random variables, the training stage of the proposed framework consists of calculating the posterior distribution of the parameters of the generative model $G$ given the training dataset $\Dcal$. At the inference stage, for a given test measurement vector ${\*y}_* \in \Cbb^M$, the proposed framework follows the BNN+LV formulation and computes the predictive distribution $p_{\mbfsf{{x}_*} | \mbfsf{{y}_*}, \mathsf{D}}  (\mbfsf{{x}_*} | {\*y}_*, \Dcal)$ by calculating the the following integral:
\begin{equation}
    % p_{\mbfsf{{x}_*} | \mbfsf{{y}_*}, \mathsf{D}}  (\mbfsf{{x}_*} | {\*y}_*, \Dcal) = \int_{\Rbb^P} \int_{\Rbb^Z} p_{\mbfsf{x} | \mbfsf{y}, \mbfsf{z}, \mbfsf{\Theta}} (\mbfsf{{x}_*} | {\*y}_*, \*z, \btheta) p_{\mbfsf{\Theta} | \mathsf{D}}(\btheta | \Dcal) p_{\mbfsf{z}}(\*z)  d \*z d \btheta, 
    \int_{\Rbb^P} \int_{\Rbb^Z} p_{\mbfsf{x} | \mbfsf{y}, \mbfsf{z}, \mbfsf{\Theta}} (\mbfsf{{x}_*} | {\*y}_*, \*z, \btheta) p_{\mbfsf{\Theta} | \mathsf{D}}(\btheta | \Dcal) p_{\mbfsf{z}}(\*z)  d \*z d \btheta, 
\label{eq:bnn-formulation-integral}
\end{equation}
where the vector $\mbfsf{\Theta}$, which is modeled as an $\Rbb^P$-valued random vector, contains the parameters of the generative model $G$ in a vectorized form; the conditional distribution $p_{\mbfsf{x} | \mbfsf{y}, \mbfsf{z}, \mbfsf{\Theta}}(\mbfsf{x} | \*y, \*z, \btheta)$ determines how the generative model $G$ with parameters $\mbfsf{\Theta}=\btheta$ maps a given measurement vector $\mbfsf{y} = \*y$ and a latent variable $\mbfsf{z}=\*z$ to the corresponding underlying image; the distribution $p_{\mbfsf{\Theta} | \mathsf{D}}(\mbfsf{\Theta} | \Dcal)$ is the posterior distribution of the parameters of the generative model given the training dataset $\mathsf{D} = \Dcal$; and $p_{\mbfsf{z}}$ is the prior distribution of the latent variable. In the rest of this subsection, we first present the design choice we have made for the form of the conditional distribution. Then, we provide the details of the procedure followed by the proposed framework to come up with a surrogate distribution for the true posterior distribution of the parameters of the generative model given the training dataset. Finally, we show how the proposed framework approximates the integral in \eqref{eq:bnn-formulation-integral} to obtain an ensemble of reconstructed images, aleatoric uncertainty estimates, and epistemic uncertainty estimates.

The proposed framework defines the conditional distribution as a specific instance of the more general conditional distribution definition provided in the BNN+LV framework~\cite{Depeweg2017BNNLV, Depeweg2018BNNLV}, as follows:
\begin{equation}
    p_{\mbfsf{x} | \mbfsf{y}, \mbfsf{z}, \mbfsf{\Theta}}(\mbfsf{x} | \*y, \*z, \btheta) = \Ncal(\mbfsf{x} | G(\*y, \*z; \btheta), \epsilon^2 \*I),
\label{eq:cond-distribution-defn}
\end{equation}
where the scalar $\epsilon > 0$ is assumed to be a fixed small constant. Although this definition might seem restrictive, it actually enables the proposed method to capture complex inherent uncertainty patterns. It is easy to verify that for fixed $\mbfsf{\Theta} = \btheta$ and $\mbfsf{y} = \*y$, the form of the conditional distribution in \eqref{eq:cond-distribution-defn} implicitly assumes that $\mbfsf{x} = G(\*y, \mbfsf{z}; \btheta) + \epsilon \mbfsf{n}$, indicating that the inherent randomness on the underlying image $\mbfsf{x}$ is modeled with the latent variable $\mbfsf{z} \sim p_{\mbfsf{z}}$ and the additive Gaussian noise $\mbfsf{n} \sim \Ncal(\*0, \*I)$. Because the generative model $G$ is capable of performing highly complex and nonlinear transformations on the latent variable $\mbfsf{z}$, the conditional distribution in \eqref{eq:cond-distribution-defn} is actually capable of representing a rich class of randomness patterns on the underlying image.

To compute the predictive distribution using \eqref{eq:bnn-formulation-integral}, we have to compute the posterior distribution of the parameters of the generative model, $p_{\mbfsf{\Theta} | \mathsf{D}}(\mbfsf{\Theta} | \Dcal)$, which corresponds to the training stage of the proposed framework. Unfortunately, calculating the exact posterior distribution of the parameters is intractable due to the deep non-linear structure of modern generative models. In Bayesian deep learning literature, numerous approaches have been suggested to address this problem for discriminative models (refer to Section \ref{sec:introduction} for specific examples). In the proposed framework, we have decided to use the deep ensembling method introduced in \cite{Lakshminarayanan2017ensembling} since it enhances the usability of the proposed framework for imaging problems. However, in principle, any suitable method may also be utilized within this framework to approximate the posterior distribution of the parameters. During the training phase, i.e., at the ensembling stage, we create $T_2$ copies of the generative model $G$ with different random initializations of the parameters and train each copy by following the training recipe $\Tcal$ on the training dataset $\Dcal$. From a probabilistic viewpoint, we can interpret this ensembling operation as an attempt to design a surrogate distribution $q$ for the true posterior distribution of the parameters $p_{\mbfsf{\Theta} | \mathsf{D}}(\mbfsf{\Theta} | \Dcal)$, where the surrogate distribution $q$ has the following form:
\begin{equation}
    q(\mbfsf{\Theta}) = \frac{1}{T_2} \sum_{t_2=1}^{T_2} \delta\left(\mbfsf{\Theta} - \tilde{\btheta}_{t_2}\right),
\label{eq:surrogate-distribution}
\end{equation}
where $\delta$ denotes the Dirac delta function, and the set $\{ \tilde{\btheta}_{t_2} \}$ contains the parameters of the trained generative models in the ensemble. It is worth noting that this ensembling procedure, hence the training stage of the proposed framework, does not require any modifications on the training procedure of the underlying generative model-based posterior sampling method $(G, p_{\mbfsf{z}}, \Tcal)$. Hence, the proposed framework is conveniently deployable for imaging problem for which we have the open-source implementation of a generative model-based posterior sampling method at hand.

Finally, at the inference stage, we approximate the predictive distribution defined in \eqref{eq:bnn-formulation-integral} by approximating the integrals with $T_1$ and $T_2$ samples stochastically and replacing the intractable posterior distribution of the parameters of the generative model, $p_{\mbfsf{\Theta} | \mathsf{D}}(\mbfsf{\Theta} | \Dcal)$, with the surrogate distribution $q$ defined in \eqref{eq:surrogate-distribution}. The resulting approximation of the predictive distribution has the following form:
\begin{equation}
   p_{\mbfsf{{x}_*} | \mbfsf{{y}_*}, \mathsf{D}}  (\mbfsf{{x}_*} | {\*y}_*, \Dcal) \approx \frac{1}{T_1 T_2} \sum_{t_1=1}^{T_1} \sum_{t_2=1}^{T_2} \Ncal(\mbfsf{{x}_*} | \bmu_{t_1, t_2}, \epsilon^2 \*I),
\label{eq:pred-dist-approx}
\end{equation}
where $\bmu_{t_1, t_2} \triangleq G({\*y}_*, \tilde{\*z}_{t_1}; \tilde{\btheta}_{t_2})$; and the set $\{ \tilde{\*z}_{t_1} \}$ contains $T_1$ samples from the prior distribution of the latent variable $p_{\mbfsf{z}}$. Since this approximation has the form of a mixture of Gaussians with uniform weights, we can easily generate samples from this distribution to obtain an ensemble of reconstructed images for the test measurement vector $\*y_*$. Moreover, we can compute the mean of this distribution using the following closed-form expression:
\begin{equation}
    \bmu = \frac{1}{T_1 T_2} \sum_{t_1=1}^{T_1} \sum_{t_2=1}^{T_2} \bmu_{t_1, t_2}.
\label{eq:pred-mean}
\end{equation}
In addition to the reconstructed images, the proposed framework is also capable of providing different types of uncertainty estimates. One way to obtain a predictive (total) uncertainty estimate is by computing the covariance matrix of this distribution, which takes the following form:
\begin{equation}
\begin{aligned}
    \bsigma_{\text{pred}} = \epsilon^2 \*I + \frac{1}{T_1 T_2} \sum_{t_1=1}^{T_1} \sum_{t_2=1}^{T_2} {\bmu}_{t_1,t_2}  			 {\bmu}_{t_1,t_2}^\top -   {\bmu}{\bmu}^\top,
\end{aligned}
\label{eq:predictive-uncertainty}
\end{equation}
where $(\cdot)^\top$ denotes the transpose operator. By following the uncertainty decomposition idea presented in \cite{Depeweg2018BNNLV}, the proposed framework can decompose the predictive uncertainty estimate in \eqref{eq:predictive-uncertainty} into epistemic and aleatoric uncertainty estimates as follows:  
\begin{align}
    \bsigma_\text{epis} & = \frac{1}{T_2} \sum_{t_2=1}^{T_2} \bar{\bmu}_{t_2}  \bar{\bmu}_{t_2}^\top - {\bmu}   {\bmu}^\top \label{eq:epistemic-uncertainty} \\
    \bsigma_\text{alea} & = \bsigma_{\text{pred}} - \bsigma_\text{epis} \label{eq:aleatoric-uncertainty}
\end{align}
where $\bar{\bmu}_{t_2} \triangleq \frac{1}{T_1} \sum_{t_1=1}^{T_1}{\bmu}_{t_1,t_2}$ for every $t_2 \in [T_2]$. Pseudo-code for the training and inference stages of the proposed framework is provided in the supplementary material.

\subsection{Comparison with Generative Model-Based Posterior Sampling Methods}
\label{ssec:proposed-method-comparison-with-gm}
To demonstrate how the proposed framework relates to the generative model-based posterior sampling method $(G, p_{\mbfsf{z}}, \Tcal)$ used within the proposed framework, let us focus on the case where the ensemble size $T_2 = 1$. If we denote the set of parameters of the trained generative model by $\btheta_*$, for a given test measurement ${\*y}_*$, the underlying generative model-based posterior sampling method generates samples from the posterior distribution of the underlying image by evaluating $G({\*y}_*, \mbfsf{z}; \btheta_*)$ for various realizations of the latent random variable $\mbfsf{z} \sim p_{\mbfsf{z}}$. Assuming that the generative model-based posterior sampling method utilizes $T_1$ realizations of the latent variable, it provides $T_1$ posterior samples $\*r_1, \dots, \*r_{T_1}$, where each posterior sample is defined as $\*r_{t_1} =  G ({\*y}_*, \tilde{\*z}_{t_1}; {\btheta}_*)$ for $t_1 \in [T_1]$, and an uncertainty estimate that is obtained by calculating the sample covariance of the posterior samples:
\begin{equation}
\begin{aligned}
    \bsigma_\text{post} = \frac{1}{T_1 - 1} \left( \sum_{t_1=1}^{T_1} {\*r}_{t_1} {\*r}_{t_1}^\top - T_1 \bar{\*r} \bar{\*r}^\top \right),
\end{aligned}
\label{eq:posterior-sampling-pred-uncertainty}
\end{equation}
where $\bar{\*r} \triangleq \frac{1}{T_1} \sum_{t_1=1}^{T_1}{\*r}_{t_1}$ is the mean of the reconstructions.

For this case, if we focus on the formulation provided by the proposed framework, the surrogate distribution takes the form of a Dirac delta function, i.e., $q(\mbfsf{\Theta}) = \delta(\mbfsf{\Theta} - \btheta_*)$, hence the approximation of the predictive distribution in \eqref{eq:pred-dist-approx} has the following form:
\begin{equation}
        p_{\mbfsf{{x}_*} | \mbfsf{{y}_*}, \mathsf{D}}  (\mbfsf{{x}_*} | {\*y}_*, \Dcal) \approx \frac{1}{T_1} \sum_{t_1=1}^{T_1} \Ncal({\*x}_* | {\*r}_{t_1}, \epsilon^2 \*I).
\label{eq:predictive-distribution-gaussian-point-parameter}
\end{equation}
It is important to highlight that each element of the set $\{ {\*r}_{t_1} \mid t_1 \in [T_1] \}$ is a sample from the posterior distribution by the assumption made by the underlying generative model-based posterior sampling method. Hence, the predictive distribution approximation in \eqref{eq:predictive-distribution-gaussian-point-parameter} can be perceived as putting point-like masses around the samples generated from the posterior distribution since $\epsilon$ is assumed to be a small constant. Moreover, for the $T_2=1$ case considered here, the predictive uncertainty information provided by the proposed framework boils down to 
\begin{equation}
\begin{aligned}
    {\bsigma}_\text{pred} &= \epsilon^2 \*I + \frac{1}{T_1} \sum_{t_1=1}^{T_1} {\*r}_{t_1} {\*r}_{t_1}^\top - \bar{\*r}  \bar{\*r}^\top.
\end{aligned}
\label{eq:proposed-point-parameter-pred-uncertainty}
\end{equation}
By comparing \eqref{eq:posterior-sampling-pred-uncertainty} and \eqref{eq:proposed-point-parameter-pred-uncertainty}, we conclude that the predictive uncertainty estimate provided by the proposed framework approaches the uncertainty estimate provided by the underlying generative model-based posterior sampling method as $\epsilon \to 0$ and $T_1 \to \infty$. As a result, we can claim that the proposed framework enhances the underlying generative model-based posterior sampling method by introducing epistemic uncertainty characterization capability without sacrificing its aleatoric uncertainty characterization capability.

\subsection{Comparison with Bayesian Neural Network-Based Image Reconstruction Methods}
\label{ssec:proposed-method-comparison-with-bnn}
The main difference between the proposed framework and the Bayesian neural network-based image reconstruction methods lies in the fundamental difference between Bayesian neural networks and Bayesian neural network with latent variables models, which involve using a latent variable $\mbfsf{z}$ in the problem formulation. More specifically, Bayesian neural network-based image reconstruction methods often assume the following form for the conditional distribution.
\begin{equation}
p_{\mbfsf{x} | \mbfsf{y}, \mbfsf{\Psi}, \mbfsf{\Phi}}(\mbfsf{x} | \*y, \bpsi, \bphi) = \Ncal(\mbfsf{x} | f(\*y; \bpsi), \Sigma(\*y; \bphi)),
\end{equation}
where the function $f: \Cbb^M \to \Cbb^N$ is a deep neural network that maps a measurement to a point on the image space; and the covariance matrix $\Sigma \in \Cbb^{N \times N}$ captures the aleatoric uncertainty, possibly through another deep neural network, i.e., $\Sigma: \Cbb^M \to \Cbb^{N \times N}$. The vectors $\mbfsf{\Psi}$ and $\mbfsf{\Phi}$, which are modeled as random vectors, contain the parameters of the neural networks $f$ and $\Sigma$ in a vectorized form, respectively. It is worth noting that this form does not include any latent variable (cf., \eqref{eq:cond-distribution-defn}) and for fixed $\mbfsf{\Psi}=\bpsi$, $\mbfsf{\Phi}=\bphi$, and $\mbfsf{y}=\*y$, it assumes that $\mbfsf{x} = f(\*y; \bpsi) + \mbfsf{n}$, where $\mbfsf{n} \sim \Ncal (\mathbf{0}, \Sigma(\*y; \bphi))$. Hence, the aleatoric uncertainty on the underlying image is modeled as additive Gaussian noise. This can be a restrictive assumption for severely ill-posed imaging problems, for which the posterior distribution of the underlying image given measurements tend to be highly multimodal. On the other hand, as we have shown in Section \ref{ssec:proposed-framework-details}, the proposed framework is capable of representing complex inherent uncertainty patterns, thanks to the latent variable used in the formulation. Therefore, the proposed framework can be interpreted as an improved version of existing Bayesian neural network-based image reconstruction methods with more advanced aleatoric uncertainty characterization capability. Experiments supporting these observations are included in the supplementary material.

\subsection{Conformalization of the Proposed Framework}
\label{ssec:conformal-prediction}
The underlying assumptions and approximations made by the proposed framework could result in biased and potentially uncalibrated predictions, offering no theoretical guarantees on the predictions. In this work, we have decided to use the split conformal prediction algorithm~\cite[Section 3.4]{Angelopoulos2024ConformalPredictionBook} to achieve such a guarantee, called the frequentist marginal coverage guarantee, on the predictions of the proposed framework. 

As described in Section \ref{ssec:proposed-framework-details}, for a given test measurement vector $\*y_*$, the proposed framework can provide a set of reconstruction candidates, a single reconstructed image, and aleatoric and epistemic uncertainty estimates. This \emph{distribution-based} approach can be perceived as an instance of a \emph{set-based} approach, where the proposed framework outputs a \emph{prediction set} of the form $\{ \bmu_{t_1, t_2} \mid t_1 \in [T_1],\; t_2 \in [T_2] \}$. The single reconstructed image, and the aleatoric and epistemic uncertainty estimates can be interpreted as the summary statistics of this prediction set calculated by the operations described in \eqref{eq:pred-mean}, \eqref{eq:aleatoric-uncertainty}, and \eqref{eq:epistemic-uncertainty}.

An alternative way to form a prediction set is to choose the reconstructions for which the corresponding value of the predictive distribution exceeds a threshold, leading to the following definition for the prediction set:
\begin{equation}
    \mathcal{C}(\*{y}_*) = \{ \*x \mid s(\*y_*, \*x) \leq \hat{q} \}
    \label{eq:prediction-set}
\end{equation}
where the real-valued \emph{score function} $s$ is defined as
\begin{equation}
    s(\*y, \*x) = - \log p_{\mbfsf{{x}_*} | \mbfsf{{y}_*}, \mathsf{D}}  ( \*x | {\*y}, \Dcal),
\end{equation}
and the scalar $\hat{q}$ is the threshold that needs to be determined based on a user-specified criterion to make the prediction set satisfy a certain property.

In this work, we have decided to focus on a particular frequentist property called marginal coverage property to achieve rigorous predictions and uncertainty estimates. For a user-defined miscoverage rate $\alpha \in [0,1]$, the prediction set $C$ is said to satisfy the marginal coverage property if 
\begin{equation}
    \mathbb{P}(\mbfsf{x}_* \in \mathcal{C}(\mbfsf{y}_*)) \geq 1 - \alpha.
    \label{eq:marginal-coverage-guarantee}
\end{equation}
A simple strategy to determine the threshold $\hat{q}$ to make the prediction set satisfy this property is to form a set containing the score function values of the training examples and choose the $(1-\alpha)-$quantile of this set. Unfortunately, although this strategy is intuitive and simple, as we will show experimentally in Section \ref{ssec:experiments-conformal}, this choice of the threshold may not provide the desired marginal coverage guarantee in \eqref{eq:marginal-coverage-guarantee}. Hence, this variant of the proposed framework, which we will refer to as the uncalibrated version of the proposed framework, may provide uncalibrated results.

Although the aforementioned strategy is not successful at determining a threshold to make the prediction set satisfy the marginal coverage property, the split conformal prediction algorithm~\cite[Section 3.4]{Angelopoulos2024ConformalPredictionBook} can achieve this by leveraging a \emph{calibration dataset} $\Dcal_\text{cal} = \{ (\*x^{[i]}, \*y^{[i]}) \mid i \in [n] \}$, which is assumed to be exchangeable and distinct from the training and test datasets. For the split conformal prediction algorithm, the desired threshold can be determined as follows:
\begin{equation}
    \hat{q} = \text{Quantile}(S_1, \dots, S_n; (1 - \alpha)(1 + 1 / n)),
    \label{eq:conformal-threshold}
\end{equation}
where the scalar $S_i \in \Rbb$ is defined as $S_i = s(\*y^{[i]}, \*x^{[i]})$. After determining the threshold $\hat{q}$, for a given test measurement vector $\*y_*$, the output of the conformalized version of the proposed framework will be the conformal prediction set $\mathcal{C}(\*y_*)$, which is a sub-level set of the negative logarithm of the predictive distribution. It has been theoretically shown that this set satisfies the marginal coverage property in \eqref{eq:marginal-coverage-guarantee} (see \cite[Section 3.4]{Angelopoulos2024ConformalPredictionBook}), thus offering rigorous predictions and predictive uncertainty estimates.

%%%%%%%%%%%%%%%%%%%%%%%%%%%%%%%%%%%%%%%%%%%%%%%%%%%%%%%%%%%%%%%%%%%%%%%%%%%%%%%%%%%%
%%%%%%%%%%% EXPERIMENTS AND RESULTS
%%%%%%%%%%%%%%%%%%%%%%%%%%%%%%%%%%%%%%%%%%%%%%%%%%%%%%%%%%%%%%%%%%%%%%%%%%%%%%%%%%%%
\section{Experiments and Results}
\label{sec:experiments-and-results}
In this section, we evaluate the proposed framework on various image recovery problems, particularly computed tomography, magnetic resonance imaging, and image inpainting. We first assess the extent to which the uncertainty estimates provided by the proposed framework align with the essential characteristics of the aleatoric and epistemic uncertainties. Then, we examine the quality of the reconstructed images and predictive uncertainty estimates obtained by the proposed framework. Next, we investigate two computationally cheaper alternatives of the ensembling procedure used within the proposed framework and discuss its advantages and disadvantages. Finally, we empirically verify whether the conformalized version of the proposed framework satisfies the marginal coverage guarantee.

\subsection{Experimental Setup}
\label{ssec:experiments-setup}
\subsubsection{CT Experiments}
For the computed tomography (CT) experiments, we obtained $11940$ $512 \times 512$ reference images from the LUNA dataset~\cite{Setio2017LunaDataset} and resized each reference image to $256 \times 256$ pixels. Then, we normalized each reference image such that the interval $[-1000, 3000]$ Hounsfield unit (HU) was mapped into the interval $[0, 1]$. We used $11220$ of those reference images for the training dataset and split the remaining ones into two parts to be used for the validation and test datasets, each containing $100$ and $620$ reference images, respectively. For each reference image in the training, validation, and test datasets, we generated the corresponding measurements by calculating its Radon transform with $72$ views (corresponding to approximately $5 \times$ dose reduction) and adding white Gaussian noise such that signal-to-noise ratio was approximately $50$ decibels.

We used a generative adversarial network-based posterior sampling method called deep posterior sampling~\cite{Adler2019CWGAN} (DPS) to build the proposed framework. To ensure reliable posterior sampling, DPS introduces a novel discriminator formulation that addresses the well-known mode collapse problem commonly observed in generative adversarial networks. At the training stage of the proposed framework, we trained $T_2=5$ copies (initialized with different random weights) of the conditional Wasserstein generative adversarial network proposed by the DPS method. We used the validation dataset to tune the hyperparameters and monitor the individual performance of each DPS instance in the ensemble. At the inference stage, for a given test measurement vector, we first calculated the filtered backprojection of the test measurement vector and then used it as an input to all generative models in the ensemble together with a sample from the prior distribution of the latent variable. We repeated this process $T_1 = 128$ times and obtained the corresponding reconstructed images and uncertainty maps.

\subsubsection{MRI Experiments}
For the magnetic resonance imaging (MRI) experiments, we obtained $41877$ $320 \times 320$ complex knee MR reference images from the fastMRI dataset~\cite{Knoll2020fastMRI, Zbontar2019fastmriopendatasetbenchmarks} and normalized each image such that the intensity values of each magnitude image lie in the interval $[0, 1]$. We used $34742$ of the reference images for the training dataset and split the remaining reference images into two sets to be used for the validation and test datasets. The validation and test datasets contain $3521$ and $3614$ reference images, respectively. For each reference image in the training, validation, and test datasets, we generated the measurements by undersampling the full k-space data from the fastMRI dataset. We drew an independent random undersampling mask for each reference image, retaining only $20\%$ of the full k-space coefficients and achieving approximately $5\times$ acceleration.

We built the proposed method on a variational autoencoder~\cite{Kingma2013VAE}-based posterior sampling method proposed in \cite{Edupuganti2021VAEMRI}, which we refer to as the Uncertainty Quantifying Variational Autoencoder for MRI (UQVAE). To ensure reliable posterior sampling, UQVAE incorporates skip connections in the decoder~\cite{Dieng2019SkipConnectionVAE} to mitigate the latent variable collapse problem commonly observed in variational autoencoders. At the training stage of the proposed framework, we trained $T_2=5$ instances of the UQVAE method, initialized with different random weights. At the inference stage, for a given test measurement vector, we first performed zero-filling and then used the result as an input to each UQVAE in the ensemble. We repeated this procedure $T_1 = 128$ times and obtained the corresponding reconstructed images and uncertainty maps.

%%%%%%%%%%%%%%%%%%%%%%%%%%%%%%%
%%%%%%%%%%% FIGURE 1
%%%%%%%%%%%%%%%%%%%%%%%%%%%%%%%
\begin{figure*}[t!]
    \centering
    \includegraphics[width=0.95\textwidth]{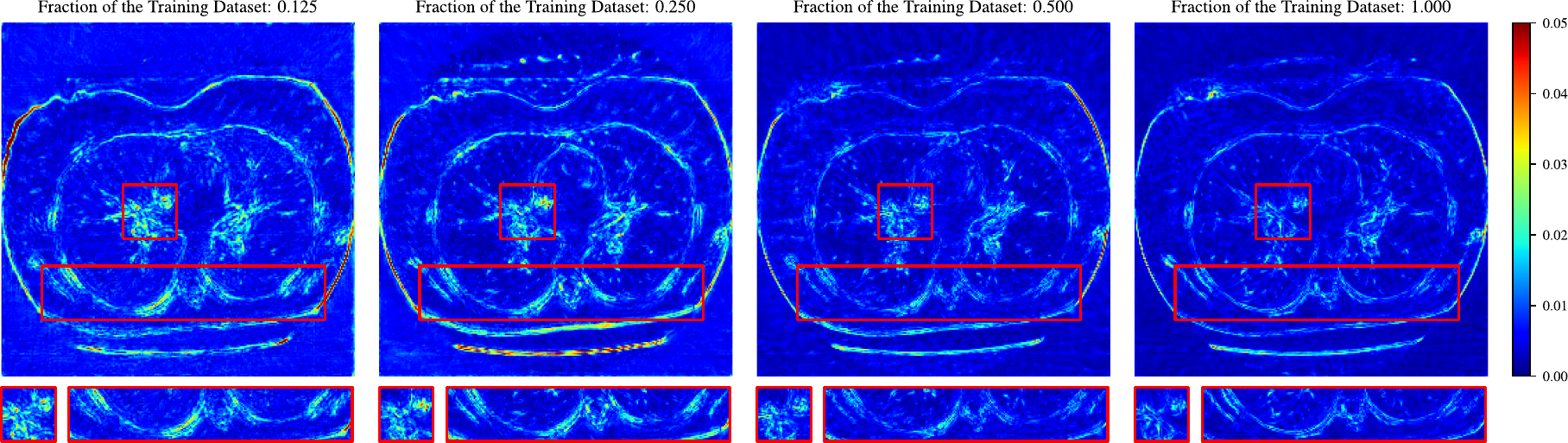}
    \caption{Variation in the epistemic uncertainty estimates offered by the proposed framework for a randomly chosen CT test measurement vector as the number of examples in the training dataset is changed. The fractions of the training data used (0.125, 0.25, 0.5, and 1.0) are indicated above each epistemic uncertainty map. For this example, the generative model-based posterior sampling method employed by the proposed framework is deep posterior sampling~\cite{Adler2019CWGAN} (see \ref{ssec:experiments-setup} for further details).}
    \label{fig:dataset-size-effect-epistemic-visual}
\end{figure*}

\subsubsection{Image Inpainting Experiments}
For the image inpainting experiments presented in Section \ref{ssec:experiments-mc-dropout}, we obtained $60025$ $32 \times 32$ reference images from the MNIST dataset~\cite{Lecun2010MNIST} and normalized them linearly such that their entries lie in the interval $[-1,1]$. We then split the reference images into two sets to be used for the training and test datasets, having $60000$ and $25$ samples, respectively. For each reference image in the test dataset, we generated its corresponding measurement vector by multiplying the image with a mask and adding white Gaussian noise. We used a mask that randomly samples $10\%$ of the image pixels and fixed the standard deviation of the noise to $0.05$.

We built the proposed framework on top of a diffusion model-based posterior sampling method called diffusion posterior sampling~\cite{Chung2023DiffusionPosteriorSampling}, whose open-source implementation is provided in \cite{Chung2022DiffusionPosteriorSamplingGithubRepo}. We trained the diffusion model~\cite{Dhariwal2021GuidedDiffusion} used within the diffusion posterior sampling technique by modifying the open-source implementation provided in \cite{Openai2021GuidedDiffusionRepo}. At the inference stage, we used each test measurement vector as an input to the diffusion posterior sampling together with $T_1 = 32$ samples from the latent distribution. We repeated this procedure $T_2 = 5$ times and obtained the corresponding reconstructions and the uncertainty maps. Further implementation details of all methods used in the experiments are provided in the supplementary material to ensure clarity and reproducibility.

\subsection{Impact of the Training Dataset Size on Epistemic Uncertainty Estimates}
% \subsection{Training Dataset Size vs. Epistemic Uncertainty Estimates}
\label{ssec:experiments-training-dataset-size-epistemic}
By the definition of epistemic uncertainty, the epistemic uncertainty on the parameters of the generative model $G$ used within the proposed framework must be reducible in the sense that increasing the size of the training dataset should lead to a decrease on the epistemic uncertainty levels. To observe if the epistemic uncertainty estimates offered by the proposed framework exhibit this reducibility feature, we examine the characteristics of the epistemic uncertainty estimates both qualitatively and quantitatively as we change the size of the training dataset.

For the sake of space, the experiments presented in this subsection focus only on the CT problem. We generated four different subsets of the original training dataset prepared for the CT reconstruction problem such that the resulting subsets contained $12.5 \%$, $25 \%$, $50 \%$, and $100 \%$ of the original training dataset. We then trained four different instances of the proposed framework on those subsets to analyze the effect of the training dataset on the epistemic uncertainty estimates. At the inference stage, we used each measurement vector in the test dataset as an input to those four instances of the proposed framework and generated the corresponding epistemic uncertainty maps.

\begin{figure}
    \centering
    \includegraphics[width=0.95\columnwidth]{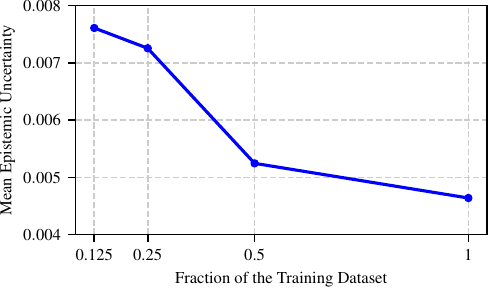}
    \caption{Average epistemic uncertainty as a function of training dataset size for the CT problem. Average values are calculated across all pixels in the test dataset. The generative model-based posterior sampling method used within the proposed framework is deep posterior sampling~\cite{Adler2019CWGAN} (see \ref{ssec:experiments-setup} for details).}
    \label{fig:dataset-size-effect-plots}
\end{figure}

Figure \ref{fig:dataset-size-effect-epistemic-visual} shows an example of four epistemic uncertainty maps obtained from a randomly chosen test measurement vector. We used the same colorbar for each map to ensure better visibility. By carefully examining the local structures of the maps, e.g., the regions indicated by the red rectangles, we see that the epistemic uncertainty decreases locally as we increase the size of the training dataset. Furthermore, by visually inspecting the maps globally, we also observe that the increase in the size of the training dataset leads to a global decrease on the epistemic uncertainty levels. We can also support that qualitative observation through quantitative analysis of the epistemic uncertainty maps. To that end, we calculated the average epistemic uncertainty per pixel over the test samples as we modified the size of the training dataset. Figure \ref{fig:dataset-size-effect-plots} depicts the resulting averages as a function of the training dataset size. As shown in the figure, we see that the overall epistemic uncertainty level decreases when we start adding more examples to the training dataset, highlighting alignment between the qualitative and the quantitative results. These observations confirm that the epistemic uncertainty estimates offered by the proposed framework exhibit the reducibility feature expected of epistemic uncertainty.

%%%%%%%%%%%%%%%%%%%%%%%%%%%%%%%
%%%%%%%%%%% FIGURE 2
%%%%%%%%%%%%%%%%%%%%%%%%%%%%%%%

%%%%%%%%%%%%%%%%%%%%%%%%%%%%%%%
%%%%%%%%%%% FIGURE 3
%%%%%%%%%%%%%%%%%%%%%%%%%%%%%%%
\begin{figure*}
    \centering
    \includegraphics[width=0.95\textwidth]{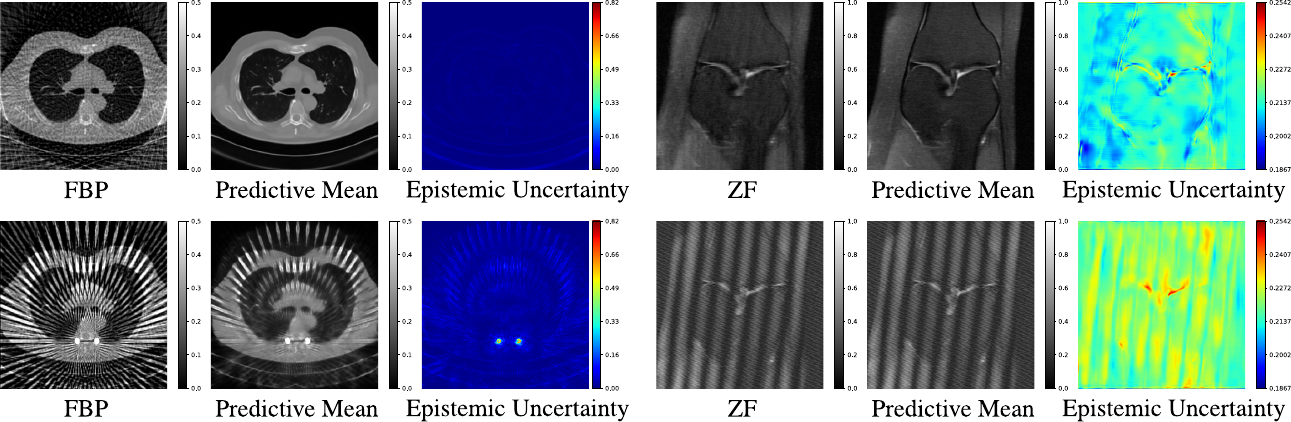}
    \caption{Effect of the test measurement vectors that are not well-represented by the training dataset on the epistemic uncertainty estimates provided by the proposed framework. The first row shows the output of filtered backprojection (FBP), the output of zero-filling (ZF), the predictive mean provided by the proposed framework, and the epistemic uncertainty estimate offered by the proposed framework for both the CT and MRI problems, where there are no abnormalities present in the test measurement vectors. The second row shows the results of the cases where abnormalities are introduced into the same test measurement vectors (please refer to Section \ref{ssec:experiments-abnormality-epistemic} for the details of these abnormalities). Note: For the MRI problem, epistemic uncertainty maps are computed before the final data consistency layer of UQVAE~\cite{Edupuganti2021VAEMRI}.}
    \label{fig:artifact-visuals}
\end{figure*}

\subsection{Epistemic Uncertainty Estimates and Abnormalities Occurring at the Inference Time}
\label{ssec:experiments-abnormality-epistemic}
The definition of epistemic uncertainty implies that the epistemic uncertainty on the parameters of the generator $G$ is caused by the absence of training examples at the vicinity of a given test measurement vector. Thus, in principle, the epistemic uncertainty must be high for a test measurement vector that is not well-represented by the training dataset. To see whether the epistemic uncertainty estimates offered by the proposed framework display this property, we intentionally introduced abnormal features, which are not well-represented by the training dataset, on test measurement vectors and examined the resulting epistemic uncertainty maps provided by the proposed framework.

For the CT problem, we inserted two synthetic metal implants on a test reference image by following the simulation procedure described in \cite{Zhang2018MetalCT, Sakamoto2019MetalCT} and generated the corresponding test measurement vector by following the procedure described in Section \ref{ssec:experiments-setup}. We then used the resulting test measurement vector as an input to the proposed framework and obtained the corresponding reconstructed image and epistemic uncertainty map. Similarly, for the MRI problem, we introduced an abnormality to a test measurement vector by adding random spikes on the Fourier transform coefficients, which is sometimes referred to as the Herringbone artifact. We then used the final test measurement vector as an input to the proposed framework.

Figure \ref{fig:artifact-visuals} displays the results for both the CT and MRI problems, as well as the results for the reference cases where no abnormalities are present on the test measurement vectors. By comparing the first and the fourth columns of Figure \ref{fig:artifact-visuals}, we see that the introduced abnormalities caused visually apparent deviations on the outputs of the FBP and ZF methods. Since the outputs of the FBP and ZF methods are essentially what the generative models used within the DPS-based proposed framework instance and the UQVAE-based proposed framework instance are conditioned upon, by examining the second and the fifth columns of Figure \ref{fig:artifact-visuals}, we observe that the reconstructed images provided by the proposed framework contain artifacts for the cases where the test measurement vectors are not well-represented by the CT and MRI training datasets. However, by looking at the epistemic uncertainty estimates shown in the third and sixth columns of Figure \ref{fig:artifact-visuals}, we see that the proposed framework has clearly identified the abnormalities on the test measurement vector as well as the artifacts caused by those abnormalities. These two examples highlight that although the proposed framework is not capable of successfully recovering the underlying image from a test measurement that is not well-represented by the training dataset, it offers a mechanism to identify and detect such problematic cases.

\subsection{Quality of the Predictive Uncertainty Estimates} 
\label{ssec:experiments-uncertainty-quality}
In this section, we assess the quality of the final uncertainty estimates offered by the proposed framework, which integrates both aleatoric and epistemic uncertainties, and compare it to the final uncertainty estimates produced by the generative model-based posterior sampling method used within the proposed framework, which quantifies only the aleatoric uncertainty. For the CT problem, we compare the proposed framework with the original DPS method. Similarly, we compare the proposed framework with the original UQVAE method for the MRI problem. Because we have $T_2=5$ different instances of the original DPS method and the UQVAE method at hand, we also compare each instance to one another to examine the variations on the quality of the uncertainty estimates induced by using different parameter values for the generative model-based posterior sampling methods. 

In our experiments, we assessed the quality of the final uncertainty estimates through the negative predictive log-likelihood metric (details provided in the supplementary material). Figure \ref{fig:quantitative-results-uncertainty} presents the negative predictive log-likelihood (NPLL) values of the evaluated methods for the CT and MRI problems. Careful examination of the figure provides several key observations about the predictive performance of the evaluated methods. First, it highlights that the quality of the predictive uncertainty estimates of both the DPS method and the UQVAE method shows notable variations depending on the initializations of the parameters. Secondly, we see that the proposed framework achieves the lowest NPLL values across all evaluated methods, demonstrating superior predictive performance compared to the DPS method and the UQVAE method. These two observations demonstrate the advantage of utilizing multiple realizations of the parameters during the inference stage, as opposed to methods that rely on a single realization of their parameters.

%%%%%%%%%%%%%%%%%%%%%%%%
% FIGURE 4
%%%%%%%%%%%%%%%%%%%%%%%%
\begin{figure}[!t]
    \centering
    % \subfloat{
    %    \includegraphics[width=0.95\linewidth]{figs/ct_mri_uncertainty_quality_bar_plots/ct_uncertainty_performance_bar_plot_nll_mixture_loss.pdf}
    %}\\
    %\subfloat{
    %    \includegraphics[width=0.95\linewidth]{figs/ct_mri_uncertainty_quality_bar_plots/mri_uncertainty_performance_bar_plot_nll_mixture_loss.pdf}
    %}
    \includegraphics[width=0.995\columnwidth]{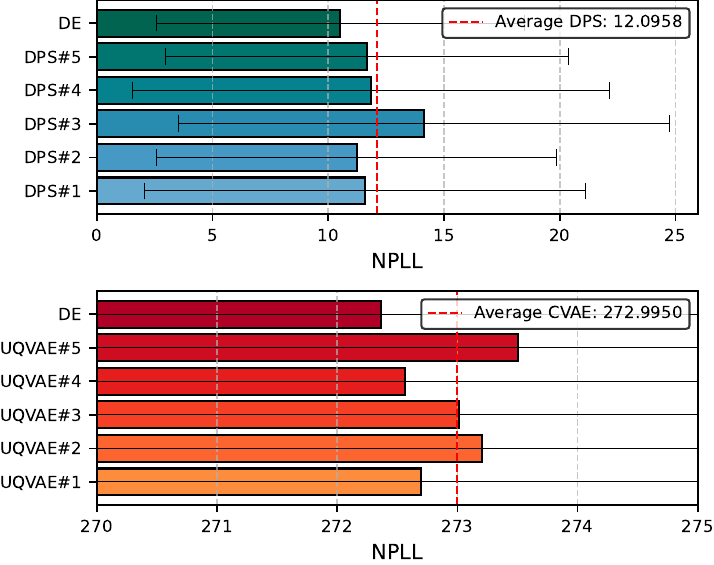}
    \caption{Negative predictive log-likelihood (NPLL) results for deep posterior sampling~\cite{Adler2019CWGAN} (DPS), the variational autoencoder-based posterior sampling method~\cite{Edupuganti2021VAEMRI} (UQVAE), and the proposed deep ensembling~\cite{Lakshminarayanan2017ensembling}-based framework (DE). The results are presented for the CT (top) and MRI (bottom) problems.}
    \label{fig:quantitative-results-uncertainty}
\end{figure}

\subsection{Reconstruction Performance} 
\label{ssec:experiments-reconstruction-quality}
The objective of this subsection is to assess the quality of the reconstructed images provided by the proposed framework and compare it against relevant baseline methods. For this purpose, we used structural similarity index~\cite{Wang2004SSIM} (SSIM) as our evaluation metric (results for the mean squared error (MSE) metric are provided in the supplementary material). When an evaluated method produced a collection of reconstructions for a given measurement vector rather than a single reconstructed image, we calculated the SSIM between the mean of the collection and the reference image.

\begin{figure}[!t]
    \centering
    % \subfloat{
    %    \includegraphics[width=0.95\linewidth]{figs/ct_mri_reconstruction_quality_bar_plots/ct_reconstruction_performance_bar_plot.pdf}
    % }\\
    %\subfloat{
    %    \includegraphics[width=0.95\linewidth]{figs/ct_mri_reconstruction_quality_bar_plots/mri_reconstruction_performance_bar_plot.pdf}
    %}
    \includegraphics[width=0.995\columnwidth]{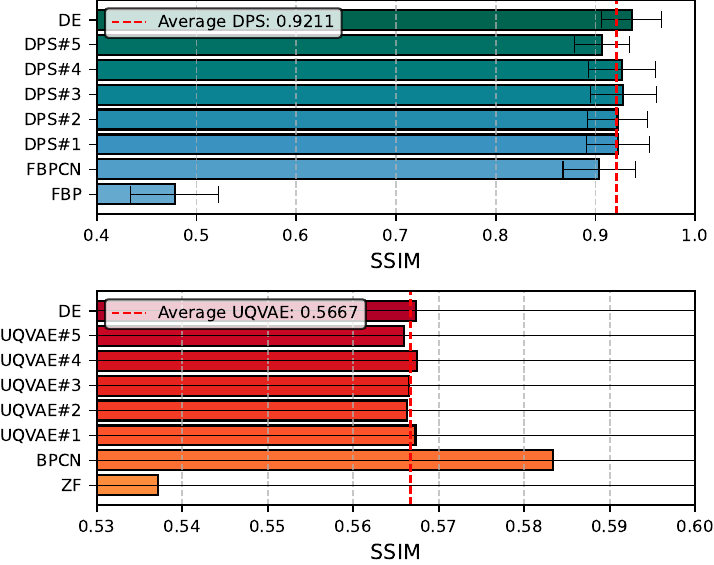}
    \caption{SSIM results for filtered backprojection (FBP), FBPConvNet~\cite{Jin2017FBPConvNet} (FBPCN), deep posterior sampling~\cite{Adler2019CWGAN} (DPS), zero-filling (ZF), BPConvNet~\cite{Jin2017FBPConvNet} (BPCN), the variational autoencoder-based posterior sampling method~\cite{Edupuganti2021VAEMRI} (UQVAE), and the proposed deep ensembling~\cite{Lakshminarayanan2017ensembling}-based framework (DE). Results shown for the CT (top) and MRI (bottom) problems.}
    \label{fig:quantitative-results-ct-mri-reconstruction}
\end{figure}

For the CT problem, we compared the performance of the proposed framework against filtered backprojection (FBP), a state-of-the-art deep learning-based image reconstruction method FBPConvNet~\cite{Jin2017FBPConvNet}, and the instances of the DPS method used within the proposed framework. For the MRI problem, we compared the performance of the proposed framework against zero-filling (ZF), a state-of-the-art deep learning-based image reconstruction method called BPConvNet~\cite{Jin2017FBPConvNet}, and the instances of the UQVAE method used within the proposed framework. Figure \ref{fig:quantitative-results-ct-mri-reconstruction} displays the resulting values of each metric calculated across the corresponding test dataset for different reconstruction methods. The visual results can be found in the supplementary material.

Through visual assessment of the reconstructed images (available in the supplementary material), we observe that the proposed framework does not lead to a significant visual improvement when contrasted with the generative model-based posterior sampling method employed within the proposed framework. Similarly, we see that initializing the same generative model-based posterior sampling method with different random parameters does not result in significant variation in the visual quality of the reconstructed images. However, the quantitative results provided in Figure \ref{fig:quantitative-results-ct-mri-reconstruction} show that the proposed framework is actually capable of improving the reconstruction performance of the inherent generative model-based posterior sampling method employed within the proposed framework, although the visual improvement may not be significant. Also, examining Figure \ref{fig:quantitative-results-ct-mri-reconstruction} further reveals that the reconstruction performance of the generative model-based posterior sampling methods is dependent on the way their parameters are initialized, demonstrating the importance of using multiple realizations of the parameters during inference for robust and improved reconstruction performance.

We also observe that, in the CT experiments, all DPS instances achieve better reconstruction performance than FBPConvNet. We believe that the superior reconstruction performance of the DPS method compared to FBPConvNet may be attributed to their different training objectives. The DPS method utilizes a loss function that aims to minimize the expected Wasserstein-1 distance between the distribution provided by the generator and the posterior distribution of the underlying image given measurements, which typically encourages the generator to output sharper and more realistic textures. On the other hand, FBPConvNet in our experiments was trained with the mean squared error loss, which is known to promote smooth reconstructions and can lead to the loss of fine details. In contrast, in the MRI experiments, we observe that BPConvNet achieves better reconstruction performance than all UQVAE instances. In our experiments, we used only the baseline configuration of the UQVAE method, i.e., without adversarial loss and without recurrent blocks. We believe this choice may explain the observed result, since those two components have been shown to provide improvements in reconstruction performance~\cite{Edupuganti2021VAEMRI}.

%%%%%%%%%%%%%%%%%%%%%%%%
% FIGURE 5
%%%%%%%%%%%%%%%%%%%%%%%%

\subsection{Computationally Efficient Ensembling}
\label{ssec:experiments-mc-dropout}
One potential drawback of the proposed framework is that training multiple instances of a generative model-based posterior sampling method could be resource intensive for certain imaging applications or particular generative model-based posterior sampling methods. The goal of this section is to investigate two computationally efficient alternatives to deep ensembling, namely MC Dropout~\cite{Gal2016MCDropout} and SWAG-Diagonal~\cite{Maddox2019SWAG}, and examine the the trade-off between computational efficiency and predictive performance.

MC Dropout, at its core, forms an ensemble of deep neural networks within \emph{a single} deep neural network by introducing dropout~\cite{Srivastava2014Dropout} layers to the network architecture and enabling them during the training and inference stages. To employ MC Dropout within the proposed framework, we have introduced dropout layers after the convolutional layers of the neural network used within the diffusion posterior sampling method. We then followed the training and inference steps described in Section \ref{ssec:experiments-setup}, with the exception that the dropout is enabled during inference. As an alternative, since the neural network architecture used within the diffusion posterior sampling method already contains dropout layers, we also tested the case where we have not added any additional dropout layers and only activated the dropout layers that are present in the original architecture. Hereafter, we refer to the first approach as MC Dropout Convolutional (MC-DC) and the second approach as MC Dropout Existing Layers (MC-DE).

Contrary to MC Dropout, SWAG does not require any changes on the training and inference procedures. It creates an ensemble by taking \emph{snapshots} of the weights during training and fitting a Gaussian distribution for each weight based on the collected snapshots. At the inference time, weight samples generated from the Gaussian distributions can be used to create an ensemble. To utilize SWAG-Diagonal within the proposed framework, we have taken snapshots during the training of the diffusion posterior sampling method, whose details are provided in Section \ref{ssec:experiments-setup}. Based on the collected snapshots, we fitted a Gaussian for each weight of the diffusion model and generated $T_2 = 5$ samples from the resulting distributions to form an ensemble. At the inference stage, we followed the steps described in Section \ref{ssec:experiments-setup}. Additional implementation details about the MC Dropout-based version of the proposed framework and the SWAG-based version of the proposed framework can be found in the source code.

\begin{figure}[!t]
    \centering
    % \subfloat{
    %    \includegraphics[width=0.95\linewidth]{figs/inpainting_reconstruction_uncertainty_quality/inpainting_reconstruction_performance_box_plot_predictive_mean.pdf}
    %}\\
    %\subfloat{
    %    \includegraphics[width=0.95\linewidth]{figs/inpainting_reconstruction_uncertainty_quality/inpainting_uncertainty_performance_bar_plot_nll_mixture_loss.pdf}
    %}
    \includegraphics[width=0.995\columnwidth]{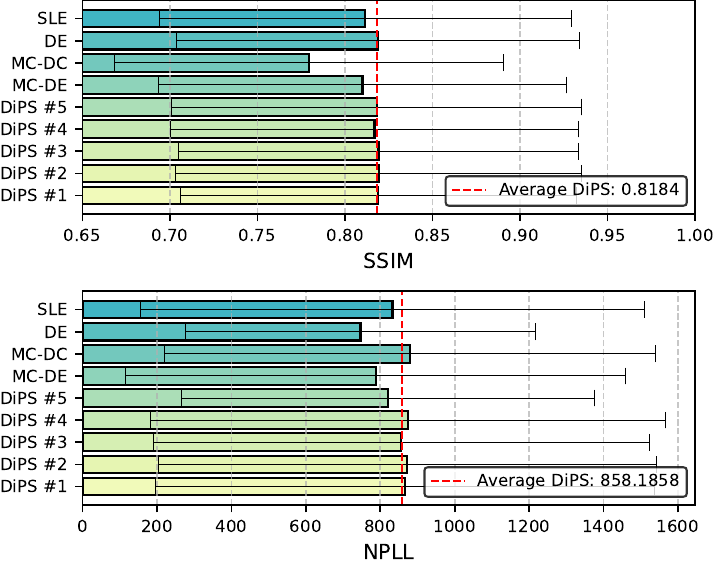}
    \caption{SSIM and negative predictive log-likelihood (NPLL) results for diffusion posterior sampling~\cite{Chung2023DiffusionPosteriorSampling} (DiPS), proposed framework with MC Dropout ensembling (MC-DC and MC-DE), deep ensembling (DE), and SWAG ensembling (SLE). The results are presented for the image inpainting problem described in Section \ref{ssec:experiments-setup}.}
    \label{fig:quantitative-results-mc-dropout}
\end{figure}

To quantitatively investigate the quality of the restored images and predictive uncertainty estimates offered by the MC Dropout- and SWAG-Diagonal-based proposed framework variants, we calculated the evaluation metrics used in Section \ref{ssec:experiments-uncertainty-quality} and Section \ref{ssec:experiments-reconstruction-quality}. Figure \ref{fig:quantitative-results-mc-dropout} displays the resulting values of the evaluation metrics computed across the test dataset. By analyzing the figure, we have identified several important insights regarding the previously mentioned computationally efficient ensembling strategies. First, as expected, we observe that deep ensembling achieves the best restoration and predictive performance compared to MC Dropout- and SWAG-based ensembling strategies at the expense of increased computational cost. Secondly, upon examination of the restoration and predictive performance of MC-DC and MC-DE, we see that the locations where dropout layers are inserted hold significance for the quality of the predictions. A straightforward application of the MC Dropout idea by adding dropout layers after convolutional layers may lead to severe restoration and predictive performance decrease. On the other hand, the use of existing dropout layers that are present in the original architecture may provide the desired improved predictive performance at the cost of a slight restoration performance decrease. Thus, we recommend conducting a through ablation study on the dropout locations rates to optimize the use of the MC Dropout-based proposed framework with a specific generative model-based posterior sampling method. Lastly, we observe that using SWAG for ensembling results in reduced restoration and predictive performance compared to deep ensembling. Nevertheless, it still outperforms individual posterior sampling instances in predictive performance, with a slight decrease in restoration performance.

%%%%%%%%%%%%%%%%%%%%%%%%
% FIGURE 5
%%%%%%%%%%%%%%%%%%%%%%%%

\subsection{Conformalization of the Proposed Framework}
\label{ssec:experiments-conformal}
\begin{figure}[!t]
    \centering
    \includegraphics[width=0.81\columnwidth]{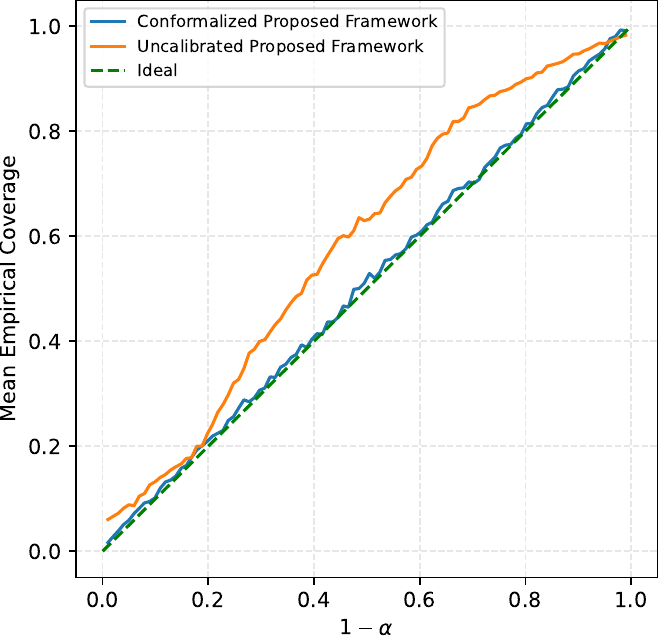}
    \caption{Mean empirical coverage versus miscoverage rate for the calibrated and uncalibrated versions of the proposed framework. The results are presented for the image inpainting problem described in Section \ref{ssec:experiments-setup}.}
    \label{fig:calibration-plot}
\end{figure}

In Section \ref{ssec:conformal-prediction}, we have provided a guideline on how to conformalize the proposed framework. In this section, we empirically test whether the prediction sets provided by the conformalized version of the proposed framework meet the desired marginal coverage guarantee described in \eqref{eq:marginal-coverage-guarantee}. For this purpose, we used the image inpainting problem as a representative image restoration problem. Since the marginal coverage performance of conformal prediction depends on the size of the calibration dataset~\cite[Theorem 4.1]{Angelopoulos2024ConformalPredictionBook}, we decided to use $200$ MNIST examples as the data, randomly splitting it into two to form our calibration and test datasets, each containing $100$ examples. For a given miscoverage rate $\alpha$, we calculated the conformal threshold $\hat{q}$ using the calibration dataset, as described in \eqref{eq:conformal-threshold}, and then formed the conformal prediction sets for each test example. To obtain an empirical estimate of the marginal coverage, we calculated the average number of test examples for which the ground image lie within the computed prediction sets. We repeated this procedure for $100$ different values of the miscoverage rate linearly spaced between $0.01$ and $0.99$ and for $100$ different random calibration-test dataset splits of the $200$ total examples. Moreover, to illustrate the impact of the calibration stage outlined in Section \ref{ssec:conformal-prediction}, we repeated a similar procedure for the  uncalibrated version of the proposed framework. The difference is that for the uncalibrated case, the threshold was determined by using a subset of the training dataset containing $100$ examples and by following the \emph{simple} threshold selection strategy described in Section \ref{ssec:conformal-prediction}. Figure \ref{fig:calibration-plot} shows the mean empirical coverage for the calibrated and the uncalibrated variants of the proposed framework as a function of the miscoverage rate.

%%%%%%%%%%%%%%%%%%%%%%%%
% FIGURE 5
%%%%%%%%%%%%%%%%%%%%%%%%

As can be seen from the figure, the prediction sets provided by the calibrated version of the proposed framework achieve nearly ideal marginal coverage. On the other hand, the uncalibrated version of the proposed framework lacks the marginal coverage guarantee and outputs underconfident predictions (larger-than-necessary prediction sets). These observations suggest that the conformalization strategy described in Section \ref{ssec:conformal-prediction} is successful at calibrating the proposed framework and ensuring accurate marginal coverage. This highlights the importance of using the conformal prediction algorithm to obtain robust predictions and reliable uncertainty estimates.

%%%%%%%%%%%%%%%%%%%%%%%%%%%%%%%%%%%%%%%%%%%%%%%%%%%%%%%%%%%%%%%%%%%%%%%%%%%%%%%%%%%%
%%%%%%%%%%% DISCUSSION
%%%%%%%%%%%%%%%%%%%%%%%%%%%%%%%%%%%%%%%%%%%%%%%%%%%%%%%%%%%%%%%%%%%%%%%%%%%%%%%%%%%%
\section{Discussion}
\label{sec:discussion}
The experimental results presented in Section \ref{sec:experiments-and-results} demonstrated the characteristics of the uncertainty estimates provided by the proposed framework and evaluated the quality of the predictive uncertainty estimates and reconstructed images. We observed that the epistemic uncertainty estimates of the proposed framework display the reducibility behavior and indicate the test cases that are not well-represented by the training dataset. Moreover, we observed that the proposed framework can improve the quality of the predictive uncertainty estimates and the reconstructed images offered by the underlying generative model-based posterior sampling method employed by the proposed framework. Furthermore, we observed that the proposed framework can be easily conformalized to form prediction sets that meet frequentist coverage guarantees, thus providing reliable and robust prediction assurances.

The aforementioned observations suggest that the proposed method has the potential to benefit various imaging applications. The reducibility feature of the epistemic uncertainty estimates provided by the proposed framework can be utilized for imaging applications for which determining the amount of training examples required to ensure confidence in generated samples is cost-intensive. Furthermore, the epistemic uncertainty estimates of the proposed framework can be leveraged to address critical challenges in safety-critical imaging applications such as out-of-distribution detection, distribution shift identification, and anomaly detection. Lastly, the epistemic uncertainty estimates provided by the proposed framework can also be used in machine learning applications such as active learning~\cite{Cohn1996ActiveLearning, Gal2017ActiveLearning} where the unlabeled inputs with high epistemic uncertainty can be forward to an oracle for labeling.

Although the proposed framework offers important benefits in various practical scenarios, it exhibits a particular limitation caused by the computational burden of ensembling. It is worth mentioning that the key motivation behind the use of deep ensembling is to make the proposed framework readily applicable for imaging problems for which there already exist open-source implementations of certain posterior sampling methods. In principle, as we have mentioned in Section \ref{ssec:proposed-framework-details}, any convenient posterior approximation technique can be utilized within the proposed framework. In Section \ref{ssec:experiments-mc-dropout}, we have briefly discussed two computationally efficient versions of the proposed framework utilizing MC Dropout and SWAG. We demonstrated that the proposed framework can still be effectively utilized for imaging applications where training multiple instances of a generative model-based posterior sampling method would be computationally expensive.

It is also important to note that the epistemic uncertainty estimates produced by the proposed framework are inherently dependent on both the size of the ensemble and the distribution of the parameters within the ensemble. If the ensemble size is small, or if the parameters of the generative models in the ensemble are concentrated around the similar modes of the posterior distribution of the parameters of the generative model given the training dataset, the resulting epistemic uncertainty estimates provided by the proposed method may suffer from bias due to limited exploration of the parameter space. In our experiments, we empirically found that an ensemble size of five was sufficient to obtain qualitatively meaningful epistemic uncertainty estimates, as well as quantitative improvements in both reconstruction and predictive uncertainty quality. Nevertheless, determining the appropriate ensemble size and developing methods to further diversify the parameter samples within an ensemble remain important questions for future research.

Finally, we emphasize that throughout this paper we have assumed that we have access to a generative model–based posterior sampling method that is expressive enough to approximate the true posterior distribution of the image given measurements. Consequently, we focused only on the uncertainty on the parameters of the generator of this posterior sampling method. However, there is also uncertainty in the choice of the generative model–based posterior sampling method itself, i.e., whether a given generative model–based posterior sampling method can adequately approximate the true posterior distribution of the image given measurements. Quantifying this uncertainty is quite challenging since it would require specifying a probability distribution over the space of generative model-based posterior sampling methods. Nonetheless, simply ignoring this uncertainty may lead to model misspecification and biased epistemic uncertainty estimates. The quantification of this uncertainty is another interesting research problem, likely requiring case-specific analysis for the imaging inverse problem of interest.

%%%%%%%%%%%%%%%%%%%%%%%%%%%%%%%%%%%%%%%%%%%%%%%%%%%%%%%%%%%%%%%%%%%%%%%%%%%%%%%%%%%%
%%%%%%%%%%% CONCLUSION
%%%%%%%%%%%%%%%%%%%%%%%%%%%%%%%%%%%%%%%%%%%%%%%%%%%%%%%%%%%%%%%%%%%%%%%%%%%%%%%%%%%%
\section{Conclusion}
\label{sec:conclusion}
In this work, we proposed a framework that is capable of quantifying aleatoric and epistemic uncertainties in imaging inverse problems. This is accomplished by incorporating existing generative model-based posterior sampling methods with Bayesian neural networks that include latent variables. We established a connection between Bayesian neural network-based image reconstruction methods and generative model-based posterior sampling methods by positioning the proposed framework with respect to those approaches. We also offered a guideline for enhancing the rigor of predictions and uncertainty estimates by applying the split conformal prediction algorithm to our framework.

We evaluated the proposed framework on several imaging problems and utilized a different generative model-based posterior sampling method to build the proposed framework for each of those problems, demonstrating the versatility of the proposed framework. The results reveal that the uncertainty estimates offered by the proposed framework display the characteristics of the true uncertainties, hence could be invaluable in practice, especially for safety-critical imaging applications. Moreover, we observed that the proposed framework is capable of improving the quality of the reconstructed images and the predictive uncertainty estimates of the underlying generative model-based posterior sampling methods. Furthermore, our results showed that applying the conformal prediction methodology on top of our approach can calibrate the proposed framework and help design prediction sets that provide frequentist coverage guarantees.

%%%%%%%%%%%%%%%%%%%%%%%%%%%%%%%%%%%%%%%%%%%%%%%%%%%%%%%%%%%%%%%%%%%%%%%%%%%%%%%%%%%%
%%%%%%%%%%% BIBLIOGRAPHY
%%%%%%%%%%%%%%%%%%%%%%%%%%%%%%%%%%%%%%%%%%%%%%%%%%%%%%%%%%%%%%%%%%%%%%%%%%%%%%%%%%%%
\bibliographystyle{IEEEtran}
\bibliography{refs}

%%%%%%%%%%%%%%%%%%%%%%%%%%%%%%%%%%%%%%%%%%%%%%%%%%%%%%%%%%%%%%%%%%%%%%%%%%%%%%%%%%%%
%%%%%%%%%%% BIOGRAPHY
%%%%%%%%%%%%%%%%%%%%%%%%%%%%%%%%%%%%%%%%%%%%%%%%%%%%%%%%%%%%%%%%%%%%%%%%%%%%%%%%%%%%
%\newpage

%\begin{IEEEbiography}[{\includegraphics[width=1in,height=1.25in,clip,keepaspectratio]{fig1}}]{Canberk Ekmekci}

%\end{IEEEbiography}

%\begin{IEEEbiography}[{\includegraphics[width=1in,height=1.25in,clip,keepaspectratio]{fig1}}]{Mujdat Cetin}

%\end{IEEEbiography}

%\vfill

\end{document}

%%%%%%%%%%%%%%%%%%%%%%%%%%%%%%%%%%%%%%%%%%%%%%%%%%%%%%%%%%%%%%%%%%%%%%%%%%%%%%%%%%%%
%%%%%%%%%%% TITLE
%%%%%%%%%%%%%%%%%%%%%%%%%%%%%%%%%%%%%%%%%%%%%%%%%%%%%%%%%%%%%%%%%%%%%%%%%%%%%%%%%%%%
\title{Supplementary Material for \\ Conformalized Generative Bayesian Imaging: An Uncertainty Quantification Framework for Computational Imaging}

%%%%%%%%%%%%%%%%%%%%%%%%%%%%%%%%%%%%%%%%%%%%%%%%%%%%%%%%%%%%%%%%%%%%%%%%%%%%%%%%%%%%
%%%%%%%%%%% AUTHOR INFO AND THANKS
%%%%%%%%%%%%%%%%%%%%%%%%%%%%%%%%%%%%%%%%%%%%%%%%%%%%%%%%%%%%%%%%%%%%%%%%%%%%%%%%%%%%
\author{Canberk Ekmekci,~\IEEEmembership{Graduate Student Member,~IEEE}, and Mujdat Cetin,~\IEEEmembership{Fellow,~IEEE}
        % <-this % stops a space
% uncomment the followings for the IEEE version of this paper
%\thanks{This work was partially supported by the National Science Foundation (NSF) under Grant CCF-1934962.}
%\thanks{Canberk Ekmekci is with the Department of Electrical and Computer Engineering, University of Rochester, Rochester, 
%NY 14627 USA (e-mail: cekmekci@ur.rochester.edu).}
%\thanks{Mujdat Cetin is with the Department of Electrical and Computer Engineering and Goergen Institute for Data Science, University of Rochester, Rochester, NY 14627 USA (e-mail: mujdat.cetin@rochester.edu).}
}

% Uncomment the following for double coulmn IEEE paper


\maketitle

\bstctlcite{IEEEexample:BSTcontrol} 

%%%%%%%%%%%%%%%%%%%%%%%%%%%%%%%%%%%%%%%%%%%%%%%%%%%%%%%%%%%%%%%%%%%%%%%%
%%%%%%%%%%% PSEUDOCODE 
%%%%%%%%%%%%%%%%%%%%%%%%%%%%%%%%%%%%%%%%%%%%%%%%%%%%%%%%%%%%%%%%%%%%%%%%
\section{Pseudo-code for Training and Inference}
\label{sec:pseudo-code}
This section provides the pseudo-code for the training and inference stages of the proposed framework. Algorithm \ref{algorithm:proposed-training} outlines the training stage of the proposed framework, which consists of forming an ensemble of optimized weights of a given generative model-based posterior sampling method (deep ensembling~\cite{Lakshminarayanan2017ensembling}). Algorithm \ref{algorithm:proposed-inference} details the steps for calculating the predictive distribution, the predictive mean, the epistemic uncertainty estimate, the aleatoric uncertainty estimate, and the predictive uncertainty estimate. All equation numbers provided in Algorithm \ref{algorithm:proposed-training} and Algorithm \ref{algorithm:proposed-inference} refer to those in the main manuscript.

%%%%%%%%%%%%%%%%%%%%%%%%%%%%%%%
%%%%%%%%%%% ALGORITHM 1
%%%%%%%%%%%%%%%%%%%%%%%%%%%%%%%
\begin{algorithm*}[t]
\caption{Training (Ensembling)}
\label{algorithm:proposed-training}
\begin{algorithmic}[1]
\State \textbf{Input: } Training dataset $\Dcal$; a generative model-based posterior sampling method $(G, p_{\mbfsf{z}}, \Tcal)$, which consists of a generative model $G$, a prior distribution $p_{\mbfsf{z}}$ on the latent variable of the generative model, and a training procedure $\Tcal$ used to train the generative model; size of the ensemble $T_2$
\For{$t_2 = 1, 2, \dots, T_2$}
\State Initialize the weights of $G$ randomly using the random seed $\text{SEED}(t_2)$.
\State Train $G$ on $\Dcal$ by following $\Tcal$ and obtain the optimized weights of the generative model $\tilde{\btheta}_{t_2}$.
\EndFor
\State \textbf{Output: }{Ensemble of optimized weights $\{\tilde{\btheta}_{t_2} \mid t_2 = 1, \dots, T_2 \}$}
\end{algorithmic}
\end{algorithm*}

%%%%%%%%%%%%%%%%%%%%%%%%%%%%%%%
%%%%%%%%%%% ALGORITHM 2
%%%%%%%%%%%%%%%%%%%%%%%%%%%%%%%
\begin{algorithm*}[t]
\caption{Inference}
\label{algorithm:proposed-inference}
\begin{algorithmic}[1]
\State \textbf{Input: } Test measurement vector ${\*y}_*$; the generative model-based posterior sampling method used at the training stage $(G, p_{\mbfsf{z}}, \Tcal)$; ensemble of optimized weights $\{\tilde{\btheta}_{t_2} \mid t_2 = 1, \dots, T_2 \}$; number of latent variable samples $T_1$
\For{$t_1 = 1, 2, \dots, T_1$}
\For{$t_2 = 1, 2, \dots, T_2$}
\State $\tilde{\*z}_{t_1} \leftarrow$ Sample a latent vector from $p_{\mbfsf{z}}$.
\State ${\bmu}_{t_1,t_2} \leftarrow$ Evaluate $G({\*y}_*, \tilde{\*z}_{t_1}; \tilde{\btheta}_{t_2})$
\EndFor
\EndFor
\State $\mathcal{E} = \{{\bmu}_{t_1,t_2} \mid t_1 = 1, \dots, T_1, \; \text{and} \; t_2 = 1, \dots, T_2 \}$  \Comment{Ensemble of reconstructed images}
\State $\bmu \leftarrow$ Compute (6). \Comment{Predictive mean (a single reconstructed image)}
\State $\bsigma_{\text{epistemic}} \leftarrow$ Compute (8). \Comment{Epistemic (generative model) uncertainty estimate}
\State $\bsigma_{\text{aleatoric}} \leftarrow$ Compute (9). \Comment{Aleatoric (posterior) uncertainty estimate}
\State $\bsigma_{\text{predictive}} \leftarrow$ Compute (7). \Comment{Predictive (total) uncertainty estimate}
\State \textbf{Output: } {$\mathcal{E}$, $\bmu$,  $\bsigma_{\text{epistemic}}$, $\bsigma_{\text{aleatoric}}$, $\bsigma_{\text{predictive}}$ }
\end{algorithmic}
\end{algorithm*}

%%%%%%%%%%%%%%%%%%%%%%%%%%%%%%%%%%%%%%%%%%%%%%%%%%%%%%%%%%%%%%%%%%%%%%%%
%%%%%%%%%%% PROPOSED VS. BNN
%%%%%%%%%%%%%%%%%%%%%%%%%%%%%%%%%%%%%%%%%%%%%%%%%%%%%%%%%%%%%%%%%%%%%%%%
\section{Comparison to Bayesian Neural Network-Based Image
Reconstruction Methods}
\label{sec:proposed-vs-bnn}
As we have mathematically shown in Section II-D, the conditional distribution assumption of the proposed framework enables capturing more complex aleatoric uncertainty patterns compared to Bayesian neural network-based image reconstruction methods, which often models the aleatoric uncertainty as an additive Gaussian noise. This section experimentally verifies this claim by generating samples from the conditional distribution of the proposed framework and the conditional distribution of a Bayesian neural network-based image reconstruction method and analyzing the resulting samples qualitatively to determine whether they behave as anticipated on a severely ill-posed inverse problem.

We decided to conduct our experiment on an image inpainting problem since both the ground truth images and the corresponding measurements lie in the image space, making the qualitative visual analysis straightforward. We utilized the MNIST dataset~\cite{Lecun2010MNIST} as in Section III-G, with the exception that the mask used in this experiment does not randomly sample $10 \%$ of the pixels; instead, it samples only the bottom half of the images. Details of the proposed framework used for this experiment are already provided in the main manuscript. For the Bayesian neural network-based image reconstruction method, we utilized the U-Net architecture~\cite{Ronneberger2015UNet} to model the mean and the covariance matrix of the conditional distribution in (13) of the main manuscript. Inspired by the strategy presented in \cite{Schlemper2018BNNUNetAndUnrolledForMRI}, our U-Net architecture has a shared downsampling path followed by two separate upsampling paths that output the mean and the diagonal entries of the covariance matrix. We utilized the MC Dropout technique~\cite{Gal2016MCDropout} to quantify the epistemic uncertainty on the parameters of the U-Net model; therefore, we trained the resulting Bayesian neural network-based image reconstruction method by minimizing a variational loss function similar to the one used in \cite{Ekmekci2022BNNImageReconstruction}. Further implementation details are provided in our codebase, which will be openly released upon the acceptance of this paper.
% At the inference stage, we generated $T_1=128$ samples from the conditional distribution and $T_2=5$ samples from the variational distribution of the parameters to obtain an approximation of the corresponding predictive distribution. 

\begin{figure}[t]
    \centering
    \subfloat{%
    \includegraphics[width=0.475\textwidth]{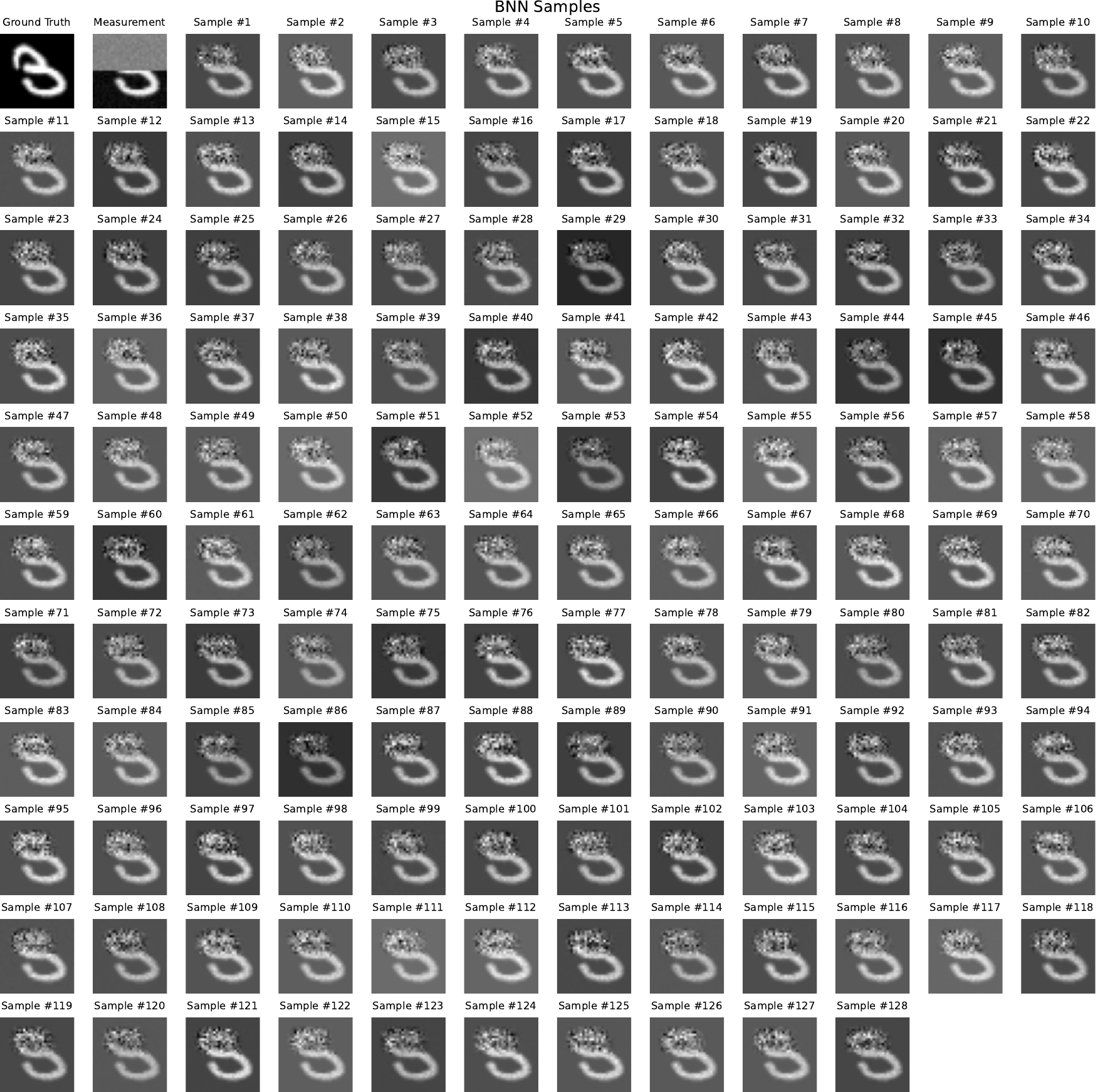}
    }
    \hfill
    \subfloat{%
    \includegraphics[width=0.475\textwidth]{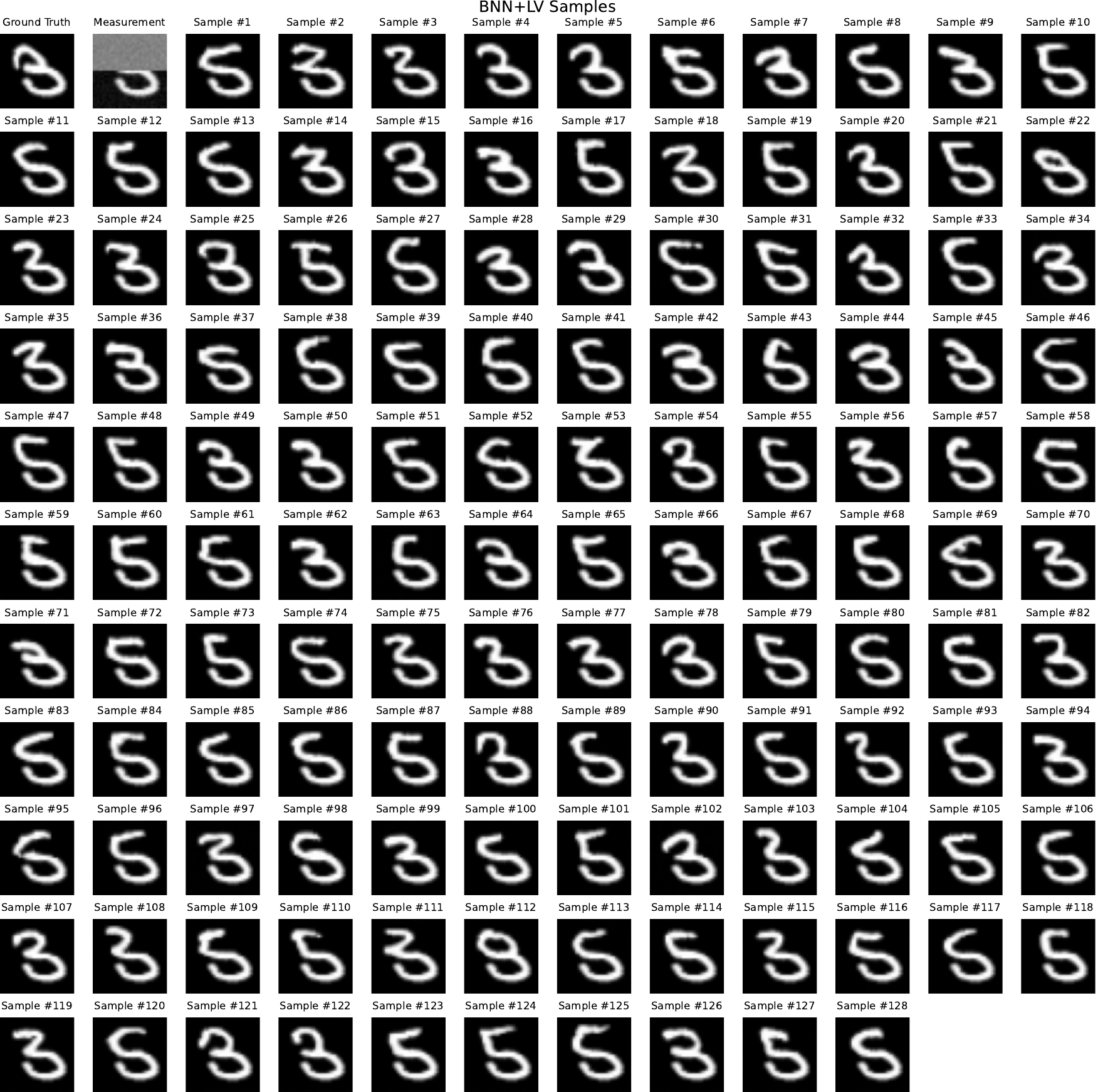}
    }
    \caption{Samples generated from the conditional distribution of a Bayesian neural network-based image reconstruction method (left) and from the conditional distribution of the proposed framework (right). For the presented severely ill-posed image inpainting problem, the conditional distribution of the proposed framework successfully captures the complex inherent aleatoric uncertainty pattern on the underlying image. On the other hand, the conditional distribution of the Bayesian neural network-based image reconstruction method is unable to do so.}
    \label{fig:bnn-vs-bnnlv}
\end{figure}

After training the proposed framework and the Bayesian neural network-based image reconstruction method, we generated $128$ samples from the conditional distribution of the proposed framework (see (3) of the main manuscript) and the conditional distribution of the Bayesian neural network-based reconstruction method (see (13) of the main manuscript). To generate samples from the conditional distribution of the proposed framework, we used one of the members of the ensemble to specify the parameter values and used $128$ realizations of the latent variable. To generate samples from the conditional distribution of the Bayesian neural network-based reconstruction method, we used one sample from the variational distribution of the parameters obtained by MC Dropout to specify the parameter values and generated $128$ samples from the resulting multivariate Gaussian distribution. Figure \ref{fig:bnn-vs-bnnlv} shows the resulting samples. 

As can be seen from the figure, the samples generated from the conditional distribution of the Bayesian neural network-based reconstruction method exhibit noise-like effects on the region to be restored. This is because the conditional distribution of the Bayesian neural network-based reconstruction method models the uncertainty on the underlying image as additive Gaussian noise. On the other hand, the proposed framework successfully restores the missing part of the image since its conditional distribution models the uncertainty on the underlying image through a deep latent generative model. These results clearly indicate that Bayesian neural network-based reconstruction methods may struggle to capture complex uncertainty patterns on the underlying images in severely ill-posed problems, whereas the proposed framework succeeds in doing so.

%%%%%%%%%%%%%%%%%%%%%%%%%%%%%%%%%%%%%%%%%%%%%%%%%%%%%%%%%%%%%%%%%%%%%%%%
%%%%%%%%%%% RECONSTRUCTION VISUALS
%%%%%%%%%%%%%%%%%%%%%%%%%%%%%%%%%%%%%%%%%%%%%%%%%%%%%%%%%%%%%%%%%%%%%%%%
\section{Visual Comparison of Reconstruction Results}
\label{sec:reconstruction-visual}
\begin{figure*}[t]
    \centering
    \includegraphics[width=0.95\textwidth]{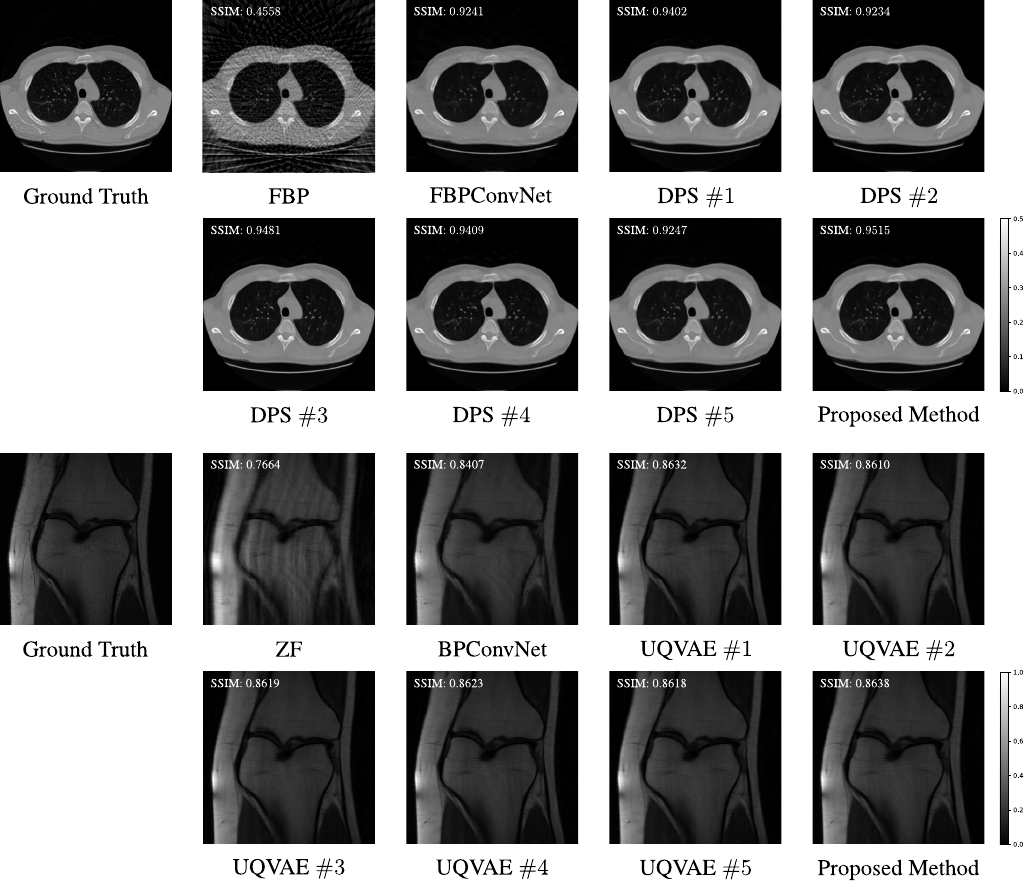}
    \caption{Reconstructed images provided by various methods for the CT and MRI problems. For the CT problem, the results of the filtered backprojection (FBP), FBPConvNet~\cite{Jin2017FBPConvNet}, five different instances of the deep posterior sampling method~\cite{Adler2019CWGAN} (DPS $\#1$-$5$) used within the proposed framework, and the proposed framework are provided. Similarly, for the MRI problem, the results of the zero filling (ZF), BPConvNet~\cite{Jin2017FBPConvNet}, five different instances of the UQVAE~\cite{Edupuganti2021VAEMRI} method (UQVAE $\#1$-$5$) used within the proposed framework, and the proposed framework are provided.}
    \label{fig:reconstructed-images}
\end{figure*}

Figure \ref{fig:reconstructed-images} presents a qualitative comparison of the reconstructed images obtained by various methods for the CT and MRI problems.  

%%%%%%%%%%%%%%%%%%%%%%%%%%%%%%%%%%%%%%%%%%%%%%%%%%%%%%%%%%%%%%%%%%%%%%%%
%%%%%%%%%%% NEGATIVE PREDICTIVE LOG-LIKELIHOOD 
%%%%%%%%%%%%%%%%%%%%%%%%%%%%%%%%%%%%%%%%%%%%%%%%%%%%%%%%%%%%%%%%%%%%%%%%
\section{Negative Predictive Log-Likelihood}
\label{sec:npll}
In the experiments presented in the main manuscript, we assessed the predictive performance of various methods using the negative predictive log-likelihood (NPLL) metric. This section provides the details about this metric, including its definition, derivation, and implementation.

The NPLL metric is defined as follows:
\begin{equation}
    \text{NPLL} = - \frac{1}{N_\text{test}} \sum_{n=1}^{N_\text{test}} \log  p_{\mbfsf{{x}_*} | \mbfsf{{y}_*}, \mathsf{D}}  \left( {\*x}_*^{[n]} \Big| {\*y}_*^{[n]}, \Dcal \right),
\end{equation}
where $N_\text{test}$ is the number of examples in the test dataset; $p_{\mbfsf{{x}_*} | \mbfsf{{y}_*}, \mathsf{D}}  ( \cdot | {\*y}_*, \Dcal )$ is the predictive distribution for the measurement vector ${\*y}_*$; and ${\*x}_*^{[n]}$ and ${\*y}_*^{[n]}$ are the ground truth image and the measurement vector corresponding to the $n^\text{th}$ test example, respectively.

For the proposed framework, by using the approximation of the predictive distribution given in (5) of the main manuscript, we obtained the following closed-form expression for the NPLL metric:
\begin{equation}
\begin{aligned}
    \text{NPLL} &= - \frac{1}{N_\text{test}} \sum_{n=1}^{N_\text{test}} \log  p_{\mbfsf{{x}_*} | \mbfsf{{y}_*}, \mathsf{D}}  \left( {\*x}_*^{[n]} \Big| {\*y}_*^{[n]}, \Dcal \right) \\
    & = - \frac{1}{N_\text{test}} \sum_{n=1}^{N_\text{test}} \log  \left( \frac{1}{T_1 T_2} \sum_{t_1=1}^{T_1} \sum_{t_2=1}^{T_2} \Ncal( {\*x}_*^{[n]} | \bmu_{t_1, t_2}^{[n]}, \epsilon^2 \*I) \right) \\
    &= - \frac{1}{N_\text{test}} \sum_{n=1}^{N_\text{test}} \log  \left( \frac{1}{T_1 T_2} \sum_{t_1=1}^{T_1} \sum_{t_2=1}^{T_2} \left[ \frac{1}{(2\pi \epsilon^2)^{N/2}} \exp\left(-\frac{1}{2\epsilon^2} \| {\*x}_*^{[n]} - \bmu_{t_1, t_2}^{[n]} \|^2 \right)   \right] \right) \\ 
    % &= - \frac{1}{N_\text{test}} \sum_{n=1}^{N_\text{test}} \log  \left( \frac{1}{T_1 T_2} \frac{1}{(2\pi \epsilon^2)^{N/2}}  \sum_{t_1=1}^{T_1} \sum_{t_2=1}^{T_2} \left[ \exp\left(-\frac{1}{2\epsilon^2} \| {\*x}_*^{[n]} - \bmu^{(t_1, t_2)}_n \|^2 \right)   \right] \right) \\
    % &= - \frac{1}{N_\text{test}} \sum_{n=1}^{N_\text{test}}  \log  \left( \frac{1}{T_1 T_2} \frac{1}{(2\pi \epsilon^2)^{N/2}} \right) - \frac{1}{N_\text{test}} \sum_{n=1}^{N_\text{test}}  \log  \left(  \sum_{t_1=1}^{T_1} \sum_{t_2=1}^{T_2} \left[ \exp\left(-\frac{1}{2\epsilon^2} \| {\*x}_*^{[n]} - \bmu^{(t_1, t_2)}_n \|^2 \right)   \right] \right) \\
    % &= \frac{1}{N_\text{test}} \sum_{n=1}^{N_\text{test}}  \log  \left( {T_1 T_2} {(2\pi \epsilon^2)^{N/2}} \right) - \frac{1}{N_\text{test}} \sum_{n=1}^{N_\text{test}}  \log  \left(  \sum_{t_1=1}^{T_1} \sum_{t_2=1}^{T_2} \left[ \exp\left(-\frac{1}{2\epsilon^2} \| {\*x}_*^{[n]} - \bmu^{(t_1, t_2)}_n \|^2 \right)   \right] \right) \\
    % &= \log  \left( {T_1 T_2} {(2\pi \epsilon^2)^{N/2}} \right) - \frac{1}{N_\text{test}} \sum_{n=1}^{N_\text{test}}  \log  \left(  \sum_{t_1=1}^{T_1} \sum_{t_2=1}^{T_2} \left[ \exp\left(-\frac{1}{2\epsilon^2} \| {\*x}_*^{[n]} - \bmu^{(t_1, t_2)}_n \|^2 \right)   \right] \right) \\
    &= \log  \left( {T_1 T_2} \right) + \frac{N}{2} \log \left( 2\pi \epsilon^2 \right) - \frac{1}{N_\text{test}} \sum_{n=1}^{N_\text{test}}  \log  \left(  \sum_{t_1=1}^{T_1} \sum_{t_2=1}^{T_2} \left[ \exp\left(-\frac{1}{2\epsilon^2} \| {\*x}_*^{[n]} - \bmu_{t_1, t_2}^{[n]} \|^2 \right) \right] \right)
\label{eq:npll-proposed}
\end{aligned}
\end{equation}
where $\bmu_{t_1, t_2}^{[n]} \triangleq G({\*y}_*^{[n]}, \tilde{\*z}_{t_1}^{[n]}; \tilde{\btheta}_{t_2})$; the set $\left\{ \tilde{\*z}_{t_1}^{[n]} \mid t_1 \in [T_1] \right\}$ contains $T_1$ samples from the prior distribution of the latent variable $p_{\mbfsf{z}}$ for the $n^\text{th}$ test example; and the set $\vartheta \triangleq \left\{ \tilde{\btheta}_{t_2} \mid t_2 \in [T_2] \right\}$ contains the parameters of the trained generative models in the ensemble. We set the scalar $\epsilon^2$ to $10^{-5}$ in our experiments and normalized the NPLL result given by \eqref{eq:npll-proposed} by the number of pixels $N$ to obtain a per-pixel metric. For generative model-based posterior sampling methods, this metric can be simply calculated by treating the ensemble set $\vartheta$ as a singleton. For Bayesian neural network-based image reconstruction methods, a similar derivation can be performed to obtain a closed-form expression for the NPLL metric.

%%%%%%%%%%%%%%%%%%%%%%%%%%%%%%%%%%%%%%%%%%%%%%%%%%%%%%%%%%%%%%%%%%%%%%%%
%%%%%%%%%%% DETAILS 
%%%%%%%%%%%%%%%%%%%%%%%%%%%%%%%%%%%%%%%%%%%%%%%%%%%%%%%%%%%%%%%%%%%%%%%%
\section{Details of the Methods Used in the Experiments}
\label{sec:model-details}
This section provides the implementation details of the methods used in the CT, MRI and inpainting experiments presented in the main manuscript.

\subsection{Deep Posterior Sampling}
For the CT reconstruction experiments presented in the main manuscript, we used a posterior sampling method called Deep Posterior Sampling (DPS)~\cite{Adler2019CWGAN}. The main reason behind this experimental choice is that the authors of this method already demonstrated the aleatoric uncertainty quantification capability of this method on the CT reconstruction problem.

DPS is based on a conditional generative adversarial network~\cite{Mirza2014ConditionalGAN} consisting of a residual U-Net~\cite{Ronneberger2015UNet} generator and a novel discriminator, which is designed to address the well-known mode collapse problem. In our CT experiments, we used the exact generator and discriminator architectures described in Appendix D.2 of \cite{Adler2019CWGAN}. We used the loss function presented in Appendix C.2 of \cite{Adler2019CWGAN} to train the generator and discriminator. We used the Adam optimizer with the learning rate of $10^{-4}$ and with the default parameters used in Pytorch. We set the mini-batch size to $16$ and performed the training for $10$ epochs.

\subsection{Uncertainty Quantifying Variational Autoencoder}
For the MRI reconstruction experiments presented in the main manuscript, we used the posterior sampling method proposed in \cite{Edupuganti2021VAEMRI}, which is specifically tailored for the MRI reconstruction problem. Consistent with the main manuscript, we refer to this method as UQVAE in this subsection. 

As its name implies, UQVAE is built upon a variational autoencoder consisting of an encoder network, a decoder network, and a data consistency layer. The encoder network takes the zero-filled reconstruction as an input and outputs a sample from the latent distribution. Then, the decoder network accepts this sample as an input and outputs a complex MR image. Finally, the resulting image is passed through a data consistency layer to generate a sample from the posterior distribution of the MR image given k-space measurements. 

The encoder and decoder network architectures we used in our MRI experiments are slightly different from those proposed in the original work~\cite{Edupuganti2021VAEMRI}. The main differences are as follows: (i) We replaced the ReLU activation functions with SiLU activation functions. (ii) We replaced each transposed convolution layer in the decoder network with bilinear upsampling followed by a convolutional layer. (iii) We replaced each batch normalization layer with a group normalization layer. (iv) We replaced the fully connected layers located at the bottleneck of the encoder and decoder networks with convolutional layers.

More specifically, the encoder network consists of 4 downsampling blocks. Each downsampling block consists of a convolutional layer followed by a group normalization layer and a SiLU activation function. The number of filters of the convolutional layers are $128$, $256$, $512$, and $1024$, respectively. The kernel size, stride, and padding of the convolutional layers are set to $5$, $2$, and $2$, respectively. Following the downsampling blocks, we have two convolutional layers modeling the mean and the logarithm of the diagonal entries of the covariance matrix of the posterior distribution of the latent variable. The number of filters and the kernel size of these convolutional layers are set to $32$ and $1$. The decoder network consists of an initial convolutional layer followed by 5 upsampling blocks. The number of filters of the initial convolutional layer is $1024$. The kernel size, stride, and padding is set to $3$, $1$, and $1$, respectively. This convolutional layer is followed by a SiLU activation function. Since the upsampling blocks are relatively more complex compared to encoder blocks, we provided a simplified pseudo-code for them in Algorithm \ref{algorithm:upsampling-block-uqvae}. For the final upsampling block, we removed the bi-cubic upsampling layer, the group normalization layer, and the SiLU activation function. For this decoder network architecture, it is worth noting that we used skip connections to attach the latent vector to the upsampling blocks as suggested by the original UQVAE method since they are shown to prevent the latent variable collapse problem for variational autoencoders~\cite{Dieng2019SkipConnectionVAE}.

We trained the encoder and decoder networks using the VAE loss function presented in Section IV of \cite{Edupuganti2021VAEMRI}. This loss function consists of an $\ell_2$ norm-based reconstruction error and a KL divergence-based regularization function. In our experiments, we replaced the $\ell_2$ norm on the reconstruction error with the $\ell_1$ norm since we experimentally observed that it led to more stable training. Furthermore, we employed a warm-up strategy on the parameter controlling the weight of the KL divergence-based regularization function, linearly increasing its value from $0$ to $1$ in the first $2$ epochs of training. We used the Adam optimizer with an initial learning rate of $10^{-4}$, which we decayed logarithmically to $10^{-5}$. We set the mini-batch size to $16$ and performed the training for $20$ epochs.

\begin{algorithm}[t]
\caption{Decoder Upsampling Block for UQVAE~\cite{Edupuganti2021VAEMRI} Used in MR Experiments}
\label{algorithm:upsampling-block-uqvae}
\begin{lstlisting}[style=pytorch]
class UpsamplingBlock(Module):
    def __init__(self, in_ch, out_ch, z_ch):
        # Bicubic upsampling
        self.ups = Upsample(scale_factor=2, mode='bicubic')
        # Main path
        self.deconv = Sequential(
            self.ups,
            Conv2d(in_ch, out_ch, kernel_size=5, padding=2, bias=False),
            GroupNorm(num_groups=min(32, out_ch), num_channels=out_ch)
        )
        # Latent projection
        self.Wz = Conv2d(z_ch, out_ch, kernel_size=1, bias=False)
        # Residual connection
        self.res_up   = self.ups
        self.res_proj = Conv2d(in_ch, out_ch, kernel_size=1, bias=False)

    def forward(self, h, z):
        # Match spatial resolution of z to h's current target resolution
        z_up = self.ups(z)
        # Main transformation
        f = self.deconv(h) 
        # Residual path: upsample + channel align if needed
        r = self.res_proj(self.res_up(h))
        # Latent contribution via 1x1 projection
        c = self.Wz(z_up) 
        # Fuse: main + latent + residual
        out = f + c + r
        # Nonlinearity
        out = SiLU(out)
        # Return fused features and the upsampled latent vector
        return out, z_up
\end{lstlisting}
\end{algorithm}

\subsection{Diffusion Posterior Sampling}
For the image inpainting experiments in the main manuscript, we used the Diffusion Posterior Sampling method~\cite{Chung2023DiffusionPosteriorSampling} (DiPS) as the generative model-based posterior sampling method. DiPS has been shown to provide diverse and high quality samples for various image restoration problems such as inpainting, deblurring, and super-resolution~\cite{Chung2023DiffusionPosteriorSampling}.

As its name implies, this method is built upon a diffusion model~\cite{Dickstein2015DiffusionModels, Ho2020DDPM, Song2019ScoreBasedModels, Song2021StochasticDifferentialEquations} modeling the prior distribution of the underlying image implicitly through a score function learned from training data. As suggested by the authors in \href{https://github.com/DPS2022/diffusion-posterior-sampling/issues/3#issuecomment-1347736532}{this response}, we used the diffusion model implementation provided in the codebase of \cite{Dhariwal2021GuidedDiffusion}, which can be found \href{https://github.com/openai/guided-diffusion?tab=readme-ov-file}{here}. We performed the training using the following parameters:

\begin{lstlisting}[language=bash, linewidth=\linewidth] 
MODEL_FLAGS="--image_size 32 --num_channels 64 --num_res_blocks 2 --attention_resolutions 16,8,4 --class_cond False --num_head_channels 32 --resblock_updown True --use_fp16 True --use_scale_shift_norm True --channel_mult 1,2,3"
DIFFUSION_FLAGS="--diffusion_steps 1000 --noise_schedule cosine --learn_sigma True"
TRAIN_FLAGS="--lr 1e-4 --batch_size 128 --save_interval 1000 --log_interval 100"
\end{lstlisting}
After performing training for $4 \times 10^{4}$ iterations, we used the open source implementation of the DiPS method provided in this \href{https://github.com/DPS2022/diffusion-posterior-sampling}{Github repository} to perform inference.

\subsection{(F)BPConvNet}
For the CT and MRI experiments, we used (F)BPConvNet as a baseline deep learning–based image reconstruction method, not for direct comparison with generative model–based posterior sampling methods, but rather to give readers a sense of how challenging the inverse problem is.

For the U-Net network~\cite{Ronneberger2015UNet} used by the (F)BPConvNet, we used the architecture described in Figure 2 of \cite{Jin2017FBPConvNet}. For the CT experiments, the input of the U-Net network is the output of the filtered backprojection operation. For the MRI experiments, the input of the U-Net network is the magnitude of the zero-filled reconstruction. We used the mean squared error loss function to train the networks. We used the Adam optimizer with the learning rate of $10^{-4}$ and with the default parameters used in Pytorch. We set the mini-batch size to $16$ and performed the training for $10$ epochs.

%%%%%%%%%%%%%%%%%%%%%%%%
% FIGURE
%%%%%%%%%%%%%%%%%%%%%%%%
\begin{figure}[!t]
    \centering
    \includegraphics[width=0.9\columnwidth]{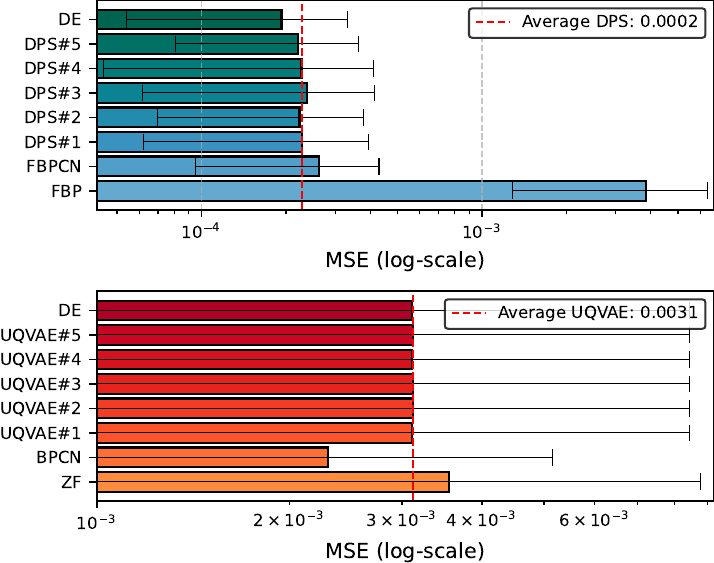}
    \caption{MSE results for filtered backprojection (FBP), FBPConvNet~\cite{Jin2017FBPConvNet} (FBPCN), deep posterior sampling~\cite{Adler2019CWGAN} (DPS), zero-filling (ZF), BPConvNet~\cite{Jin2017FBPConvNet} (BPCN), the variational autoencoder-based posterior sampling method~\cite{Edupuganti2021VAEMRI} (UQVAE), and the proposed deep ensembling~\cite{Lakshminarayanan2017ensembling}-based framework (DE). Results shown for the CT (top) and MRI (bottom) problems.}
    \label{fig:quantitative-results-ct-mri-reconstruction}
\end{figure}

%%%%%%%%%%%%%%%%%%%%%%%%%%%%%%%%%%%%%%%%%%%%%%%%%%%%%%%%%%%%%%%%%%%%%%%%%%%%%%%%%%%%
%%%%%%%%%%% BIBLIOGRAPHY
%%%%%%%%%%%%%%%%%%%%%%%%%%%%%%%%%%%%%%%%%%%%%%%%%%%%%%%%%%%%%%%%%%%%%%%%%%%%%%%%%%%%
\bibliographystyle{IEEEtran}
\bibliography{supp_refs}